\documentclass[aps, prx, reprint, longbibliography, nofootinbib,superscriptaddress, floatfix]{revtex4-1}

\usepackage{slashed}
\usepackage{verbatim}
\usepackage[T1]{fontenc}
\usepackage{mathbbol}
\usepackage[dvipsnames]{xcolor}
\usepackage{orcidlink}

\usepackage[normalem]{ulem}
\usepackage[english]{babel}
\usepackage{physics}
\usepackage{graphicx,color,overpic,mathtools}
\usepackage{amsthm,amsmath,amssymb,mathrsfs}
\usepackage{braket,bm,bbm,setspace}
\usepackage{cancel}
\usepackage{float}
\usepackage{xargs}
\definecolor{myred}{RGB}{179, 27, 27}
\usepackage{hyperref}
\hypersetup{
    colorlinks=true,
    linkcolor=myred, 
    citecolor=myred, 
    urlcolor=myred  
 }

\newcommand{\ed}{\mathop{}\!\mathrm{d}}
\newcommand{\viscgs}{\mathop{}\!\mathrm{g}/\mathrm{cm}/\mathrm{s}}
\newcommand{\epscgs}{\mathop{}\!\mathrm{g}/\mathrm{cm}^{3}}

\begin{document}

\title{Radial Oscillations of Viscous Stars}

\author{Lennox S. Keeble\,\orcidlink{0009-0009-5796-631X}}
\email{lsk35@cam.ac.uk}
\affiliation{Department of Applied Mathematics and Theoretical Physics,
University of Cambridge, Wilberforce Road, Cambridge CB3 0WA, United Kingdom}

\author{Jaime Redondo-Yuste\,\orcidlink{0000-0003-3697-0319}}
\email[]{jredondo@jhu.edu}
\affiliation{Center of Gravity, Niels Bohr Institute, Blegdamsvej 17, 2100 Copenhagen, Denmark}
\affiliation{William H. Miller III Department of Physics \& Astronomy, Johns Hopkins University\\
3400 North Charles Street, Baltimore, MD 21218, USA}

\begin{abstract}
Oscillation modes of neutron stars, a key target for third-generation gravitational wave detectors, encode key information about their constituent nuclear matter. In this work, we study the effect of viscosity on oscillations of cold, polytropic, spherically symmetric neutron stars. We focus on purely radial oscillations and work perturbatively to linear order within two hydrodynamic frameworks: the acausal covariant generalization of the Navier-Stokes equations proposed by Eckart, and the causal generalization formulated by Bemfica, Disconzi, Noronha, and Kovtun (BDNK). We find that viscosity damps the radial modes on millisecond timescales and induces fractional shifts in the oscillation frequency which increase both with the compactness and viscosity of the star, reaching up to the percent level for the fundamental mode with bulk viscosities $\zeta\sim10^{30}\viscgs$. For more viscous stars, the oscillation frequency decreases, becoming zero (i.e., an overdamped mode) for $\zeta\gtrsim10^{31}\viscgs$. We also study the linear threshold of gravitational collapse. Consistent with recent analytic results in the zero heat conductivity limit, we find that viscosity in Eckart theory cannot stabilize an unstable inviscid star. We provide numerical evidence that viscosity in BDNK theory is similarly unable to prevent gravitational collapse, but it slightly modifies the threshold of collapse. Overall, our results advance our understanding of the impact of viscosity on the oscillation modes of neutron stars, a key component of viscous asteroseismology with next-generation gravitational wave detectors.
\end{abstract}

\maketitle

\section{Introduction}

Neutron stars are a fascinating laboratory for new physics, since the four fundamental interactions of nature play a major role in their dynamics. One key puzzle is related to the composition of matter in the interior of neutron stars, where the density is most extreme~\cite{Baym:2017whm, Lattimer:2015nhk}. The conditions met in such stellar interiors (supranuclear densities at $\mathrm{keV}$ temperatures) are unlike anything we can probe on Earth, where such extreme densities can only be achieved by simultaneously probing extreme energy scales in heavy-ion collisions~\cite{Busza:2018rrf, Alford:2007xm}. Therefore, despite abundant observations of neutron stars in the Universe, we cannot confidently conclude what the composition of matter in their interior is. Does it consist of a quark-gluon plasma, a color superconductor, hyperons, or other heavy baryons?

Gravitational wave observations of neutron stars merging seemingly offer a promising avenue to answer this question. Early events such as GW170817~\cite{LIGOScientific:2017vwq, LIGOScientific:2018hze} have already provided novel constraints on the equation of state of nuclear matter. Third-generation detectors such as the Einstein Telescope and Cosmic Explorer will likely observe most of the neutron star binaries merging in the Universe, with greatly improved accuracy compared to current observations~\cite{ET:2019dnz, ET:2025xjr, Reitze:2019iox}. 

A particularly promising avenue for inferring properties of matter in the interior of neutron stars is gravitational wave asteroseismology. This program aims to deduce the composition of stars by characterizing their oscillation modes. Neutron stars possess several oscillation channels, including the usual pressure $p$-modes, and crust $g$-modes, as well as spacetime $w$-modes, interface $i$-modes associated with phase transitions, shear $s$-modes due to the elasticity of the crust, etc. A practical rule of thumb is that each layer of additional complexity in the stellar model typically contributes an additional family of modes. Stellar oscillation modes leave an imprint in the gravitational wave signal during the late inspiral (as tidal resonances), and in the post-merger phase. We refer the reader interested in their observational prospects to Refs.~\cite{Andersson:1997rn, Kokkotas:1999bd, Benhar:2004xg} and references therein.

The impact of viscosity on neutron star oscillations has received limited attention~\cite{cutler1987effect, cutler1990damping, Sawyer:1989dp}. One possible reason is that neutron stars were thought to be relatively inviscid, making the perfect fluid framework a sensible approximation. However, recent research has demonstrated that if neutron stars feature \emph{strange} quarks in their core -- either isolated, or forming \emph{hyperons} -- the situation may be quite different, and the star's bulk viscosity cannot be safely neglected~\cite{Alford:2017rxf, Alford:2018lhf, Alford:2020lla, Most:2021zvc, Most:2022yhe, Ghosh:2023vrx, Ghosh:2025glz, Ghosh:2025wfx}. In fact, the bulk viscosity can be as large as $\zeta \sim 10^{28} - 10^{30} \viscgs$ during the late inspiral and the post-merger. An observational detection of viscosity becomes a smoking gun of strangeness in the core of neutron stars. This has motivated recent activity related to understanding the impact of viscosity in the GW emission of neutron stars through tidal effects~\cite{Saketh:2024juq, Ripley:2023lsq, Ripley:2023qxo, HegadeKR:2024agt} and their oscillation modes~\cite{Redondo-Yuste:2024vdb, Boyanov:2024jge, Redondo-Yuste:2025ktt}, as well as numerical simulations of viscous neutron star mergers within M\"uller-Israel-Stewart (MIS) hydrodynamics, showing visible differences in the gravitational wave emission~\cite{Chabanov:2023abq, Chabanov:2023blf, Chabanov:2024yqv}.

Our work studies the \emph{radial} modes of viscous stars~\cite{Chanmugam:1977, Glass:1983, Kokkotas:2000up}. For inviscid stars, these have real frequencies, since there is no damping mechanism besides nonlinear coupling to non-radial modes~\cite{Passamonti:2005cz, Passamonti:2007tm}. Increasing the compactness of the star eventually results in an imaginary radial oscillation frequency -- an unstable mode -- signaling the instability of the star against gravitational collapse~\cite{Chandrasekhar_1964, Bardeen:1966}. The inclusion of viscous effects provides a dissipation channel that renders the problem of finding the radial modes of a star non-hermitian, and endows the corresponding eigenvalues with an imaginary part. 

Recently, several works have begun investigating the radial oscillations of viscous stars. The radial oscillation modes of stellar configurations within Eckart and MIS hydrodynamics were studied in Ref.~\cite{Mendes:2025oib}. The treatment is simplified in the Eckart frame, but the underlying hydrodynamic theory is ill-posed and acausal. The pathologies in the Eckart frame are cured in MIS theory, but at the expense of including additional dynamical dissipative degrees of freedom at second-order in the gradient expansion. A well-posed, causal, and stable first-order theory of dissipative hydrodynamics was formulated by Bemfica, Disconzi, Noronha, and Kovtun (BDNK)~\cite{Bemfica:2017wps, Bemfica_2018, Bemfica:2019knx, Bemfica:2020zjp,Kovtun_2019}. First steps towards nonlinear studies of radial oscillations of stars within BDNK hydrodynamics were taken in Ref.~\cite{Shum:2025jnl} in the Cowling approximation. Analytic results regarding stability criteria within Eckart, BDNK and MIS hydrodynamics in the small viscosity and perturbative regime were established in Ref.~\cite{Caballero_2025}, including, in particular, mode stability in Eckart hydrodynamics for arbitrary viscosity in the zero heat conductivity limit.

In this work, we numerically study the radial oscillations of cold, viscous, polytropic neutron stars within Eckart and BDNK hydrodynamics, working to linear order in the perturbative regime. The key questions we investigate are how viscosity (i) parametrically shifts the radial modes of a stable inviscid star, and (ii) affects the threshold and rate of gravitational collapse. Our code is written in the \texttt{Julia} programming language and is publicly available in the package \texttt{NeutronStarOscillations.jl}~\cite{Keeble_2026}.

We find that viscosity shifts the oscillation frequency of the fundamental mode downwards by up to the percent-level for compact stars with bulk viscosities $\zeta\lesssim10^{30}\viscgs$. For more viscous stars, the oscillation frequency can be arbitrarily small, vanishing for bulk viscosities $\zeta\gtrsim10^{31}\viscgs$ across two equations of state and several compactnesses.

Consistent with the stability analysis of Ref.~\cite{Caballero_2025} in the zero heat conductivity limit, our results suggest viscosity within Eckart hydrodynamics cannot stabilize an unstable inviscid star, but it can slow down the rate of collapse by several orders of magnitude in sufficiently viscous stars. In particular, we find that (very large) bulk viscosities $\zeta\gtrsim10^{33}\viscgs$ can slow down the timescale of collapse from milliseconds to seconds. In contrast to the Eckart frame, we find that viscosity within BDNK theory \emph{does} slightly modify the threshold of collapse. However, for the specific polytropic equation of state considered, viscosity in BDNK theory is unable to stabilize an unstable inviscid star, as in the Eckart frame.

The remainder of this paper is organized as follows. In Sec.~\ref{sec:EinsteinEquations}, we describe the spherically symmetric spacetime  and hydrodynamic stress-energy tensors considered in this work. We formulate the linearized equations of motion governing radial perturbations of equilibrium configurations in Sec.~\ref{sec:LinearProblem}. The numerical methods employed to solve the equations of motion are described in Sec.~\ref{sec:Numerics}. Our results are presented in Sec.~\ref{sec:Results} and our concluding discussion in Sec.~\ref{sec:Discussion}.

Throughout, we reserve letters early in the Latin alphabet for abstract index notation and work in geometric units $G=c=1$.

\section{Einstein-Navier-Stokes system in spherical symmetry}\label{sec:EinsteinEquations}

We study the coupled dynamics of a spherically symmetric spacetime and a viscous, relativistic fluid. In the area gauge, the metric is determined by two functions $\nu(t,r)$ and $\lambda(t,r)$:
\begin{equation}
    \ed{s}^2 = -e^{\nu(t,r)}\ed t^2+e^{\lambda(t,r)}\ed r^2+r^2\ed \Omega^2 \, \label{eq:metric}, 
\end{equation}
where $\ed\Omega^2$ is the unit round metric on $S^{2}$. We consider a cold, barotropic fluid whose equation of state is of the form $p=p(\epsilon)$, where $p$ and $\epsilon$ denote the fluid pressure and local energy density, respectively. The details of the fluid are encoded in the stress-energy tensor $T_{ab}= T_{ab}(\epsilon, u^c)$, where $u^{c}$ denotes the four-velocity of the fluid. The fluid is thus fully specified by $\epsilon(t,r)$ and $u^r$, with $u^{t}$ fixed by the normalization $u_au^a=-1$. Notice that, within spherical symmetry, no polar or azimuthal motion takes place. 

The dynamics of the spacetime and the fluid are coupled through the Einstein equations $G_{ab}=8\pi T_{ab}$. In a spacetime with metric \eqref{eq:metric}, they take the form
\begin{subequations}
\begin{align}
        8\pi r^2 T_{t}{}^{t}&=\partial_{r}\left(r e^{-\lambda}\right)-1,\label{eq:EFEtt}\\
        8\pi r^2 T_{r}{}^{r}&=e^{-\lambda}\left(r\partial_{r}\nu+1\right)-1,\label{eq:EFErr}\\
        32\pi T_{\theta}{}^{\theta}&=e^{-\lambda}\left[2\partial^{2}_{r}\nu-\partial_{r}\lambda\,\partial_{r}\nu+(\partial_{r}\nu)^2+\frac{2}{r}\partial_{r}\left(\nu-\lambda\right)\right]\nonumber\\
        &\quad-e^{-\nu}\left[-\partial_{t}\lambda\,\partial_{t}\nu+2\partial^{2}_{t}\lambda+(\partial_{t}\lambda)^2\right],\label{eq:EFEthth}\\
        8\pi  rT_{t}{}^{r}&=e^{-\lambda}\partial_{t}\lambda.\label{eq:EFEtr} 
\end{align}\label{eq:EFEs}
\end{subequations}
It follows from the contracted Bianchi identity that the Einstein equations \eqref{eq:EFEs} imply the conservation of the stress-energy tensor, $\nabla_aT^{ab}=0$. 

Specifying the form of the stress-energy tensor selects a theory of relativistic hydrodynamics. Hydrodynamics is an effective field theory~\cite{Kovtun:2012rj} in which we integrate out the physical degrees of freedom smaller than a certain length scale $\ell_{\rm mfp}$  (typically the microphysical mean free path). In practice, this means the stress-energy tensor is an expansion in gradients of the fundamental thermodynamic variables such as the temperature (or energy density) and fluid velocity, i.e., $T_{ab}=T^{(0)}_{ab} + T^{(1)}_{ab} + \mathcal{O}(\partial^2)$. Lorentz covariance and symmetry arguments uniquely fix the zeroth order piece,
\begin{equation}
    T^{(0)}_{ab}=\epsilon u_au_b + p\Delta_{ab} \, , 
\end{equation}
where $\Delta_{ab}=g_{ab}+u_au_b$ is the projector orthogonal to the fluid's worldline. The expansion truncated at this order, $T_{ab}=T^{(0)}_{ab}$, describes an inviscid,  ``perfect'' fluid whose entropy current is conserved for smooth solutions to the associated equations of motion~\cite{Kovtun:2012rj}. Dissipative effects such as viscosity appear only when including higher-order terms in the gradient expansion. In this work, we focus on relativistic generalizations of Navier-Stokes hydrodynamics, and hence restrict to theories which are first order in gradients. The dissipative effects are thus encoded in the first-order piece $T^{(1)}_{ab}$, and the stress-energy tensor assumes the form $T_{ab}=T^{(0)}_{ab}+T^{(1)}_{ab}$.

There is no unique prescription for $T^{(1)}_{ab}$. In general, it takes the form 
\begin{equation}
    T^{(1)}_{ab} = \mathcal{A} u_au_b + \Pi \Delta_{ab}  + 2u_{(a}\mathcal{Q}_{b)} -2\eta\sigma_{ab} \, , 
\end{equation}
where $\eta$ is the shear viscosity, $\mathcal{A}, \Pi, \mathcal{Q}_a$ are functions of gradients of $\epsilon, u^a$, the heat flux is transverse to the fluid's worldline, $\mathcal{Q}_au^a=0$, and the shear tensor is defined as 
\begin{equation}
    \sigma^{ab}=\frac{1}{2}\Delta^{ac}\Delta^{bd}\left(\nabla_{c}u_{d}+\nabla_{d}u_{c}-\frac{2}{3}\Delta_{cd}\nabla_{e}u^{e}\right).
\end{equation}
Specifying a particular dependence of the functions $\mathcal{A}(\nabla\epsilon, \nabla u),\Pi(\nabla\epsilon, \nabla u),\mathcal{Q}^a(\nabla\epsilon, \nabla u)$ constitutes a particular choice of \emph{hydrodynamic frame}. Different frames are related to each other by redefinitions of the fundamental fields of the effective theory. Perhaps the simplest frame to consider is the Landau frame~\cite{Landau:1987}, where simply 
\begin{equation}\label{eq:Landau}
    \mathcal{A}^{\rm Landau} = \mathcal{Q}_{\rm Landau}^a=0 \, , \qquad \Pi^{\rm Landau} = -\zeta \nabla_a u^a \, ,
\end{equation}
with $\zeta$ the bulk viscosity. The Landau frame coincides with the frame introduced by Eckart~\cite{Eckart:1940} in the zero diffusion limit, which is the setting of this work. For consistency with recent works \cite{Caballero_2025, Mendes:2025oib}, we henceforth refer to the frame choice \eqref{eq:Landau} as the Eckart frame. 

The Eckart frame is appealing for its simplicity, but the resulting equations of motion are known to lead to acausal propagation of signals~\cite{Cattaneo:1958, Kranys:1966} and unphysical instabilities~\cite{Hiscock:1983, Hiscock:1985}, rendering them ill-posed. Consider now a frame transformation generated by the field redefinition
\begin{equation}
    \begin{aligned}
        \epsilon \to\,& \epsilon + \frac{\tau_p}{c_{s}^{2}}\Bigl[u^a\nabla_a \epsilon + (\epsilon+p)\nabla_au^a\Bigr] \, , \\
        u^a \to\,& u^a + \frac{\tau_e-\tau_p / c_{s}^{2}}{2(\epsilon+p)}\Bigl[u^b\nabla_b \epsilon + (\epsilon+p)\nabla_bu^b\Bigr]u^a \\
        &+\tau_q \Bigl[u^b\nabla_b u^a + \frac{c_s^2}{\epsilon+p}\Delta^{ab}\nabla_b \epsilon\Bigr] \, , 
    \end{aligned}\label{eq:FieldRedefinition}
\end{equation}
which is parametrized by three transport coefficients $\tau_e,\tau_p,\tau_q$, each of which have units of time and are formally interpreted as relaxation times. Above, $c_s^2=\ed p/\ed\epsilon$ is the sound speed. On-shell (i.e., evaluated on the equations of motion), the frame transformation \eqref{eq:FieldRedefinition} satisfies $\epsilon\to\epsilon + \mathscr{O}(\partial^2)$ and $u^a\to u^a + \mathscr{O}(\partial^2)$. It is hence of type II according to~\cite{Bea:2025eov}, and the physical content of the theory should not depend on the choice of relaxation times $\tau_i$. Under this transformation, the stress-energy tensor becomes
\begin{equation}
    \begin{aligned}
        \mathcal{A}^{\rm BDNK}&=\tau_{e}\left[u^a\nabla_a\epsilon+\left(\epsilon+p\right)\nabla_au^a\right],\\
        \Pi^{\rm BDNK}&=-\zeta\nabla_{a}u^{a}+\tau_{p}\left[u^a\nabla_a\epsilon+\left(\epsilon+p\right)\nabla_au^a\right],\\
        \mathcal{Q}_{\rm BDNK}^a&=\tau_q\Bigl[\left(\epsilon+p\right)u^b\nabla_bu^a+c_s^2\Delta^{ab}\nabla_{b}\epsilon\Bigr].
    \end{aligned}\label{eq:BDNKTheory}
\end{equation}
This stress-energy tensor, manifestly more complicated than in the Eckart frame, leads to well-posed equations of motion for which perturbations \emph{always} propagate causally, provided the three relaxation times $(\tau_e,\tau_p,\tau_q)$ satisfy certain inequalities. The unphysical instabilities and acausality which plague the Eckart and Landau relativistic generalizations of Navier-Stokes theory result from a poor choice of hydrodynamic frame. Causality, stability, and well-posedness of the generalization of the relativistic Navier-Stokes theory in \eqref{eq:BDNKTheory} was proved by Bemfica, Disconzi, Noronha~\cite{Bemfica:2017wps, Bemfica_2018, Bemfica:2019knx, Bemfica:2020zjp}, and Kovtun~\cite{Kovtun_2019} --- we refer to these frames as \emph{BDNK frames}. 

We parametrize the transport coefficients in terms of (hatted) dimensionless quantities,
\begin{gather}
    \begin{aligned}
    \eta = \hat{\eta}(\epsilon+p) Lc_{s}^{2},\quad&\zeta =\hat{\zeta}(\epsilon+p) L c_{s}^{2}\\
    \tau_{e}= \hat{\tau}_{e}L\hat{V},\quad\tau_{q}=\hat{\tau}_{q}&L\hat{V},\quad \tau_{p}=\hat{\tau}_{p} L\hat{V} c_{s}^2,
    \end{aligned}\label{eq:BDNK:Frame}
\end{gather}
where $L$ is some length scale and $\hat{V}=\hat{\zeta}+4\hat{\eta}/3$. The dimensionless quantities $\hat{\zeta}$ and $\hat{\eta}$ control the amount of bulk and shear viscosity in the fluid, respectively, and thus changing their values changes the physical content of the theory. In contrast, the relaxation times $\tau_{i}$ act as causal regulators in BDNK theory; changing the dimensionless quantities $\hat{\tau}_i$ should not significantly affect the physical properties of the fluid, provided it remains within the regime of validity of the first-order truncation of the effective field theory. We monitor whether the fluid remains within the regime of validity of the theory by checking ``frame robustness''\footnote{See Ref.~\cite{Bea:2025eov} for a validation of frame robustness in the context of quark-gluon plasma, reproducing experimental results with the relativistic Navier-Stokes framework also considered here.}, i.e., that different choices of $\hat{\tau}_{i}$ change our results by an amount that is significantly smaller than the difference between the dissipative and non-dissipative hydrodynamic theories. 

To ensure causal propagation of linear perturbations, the dimensionless constants parameterizing the class of frames \eqref{eq:BDNK:Frame} must satisfy certain inequalities. By analyzing the characteristics of the Einstein-BDNK system (described in detail in Sec.~\ref{sssec:LinearProblem:CausalFrame}), we find that linear perturbations propagate causally provided $\eta,\zeta\geq 0$, $\tau_e,\tau_p,\tau_q>0$, and the following two inequalities are satisfied
\begin{equation}
    \hat{\tau}_p> 1 \, , \quad c_s^2 \leq c^{2}_{\mathrm{max}}\equiv\frac{\hat{\tau} _{e } \hat{\tau}_q}{\hat{\tau}_{e }+\hat{\tau}_q \left(\hat{\tau}_{p}+\hat{\tau}_{e }\right)} \,.\label{eq:Causality}
\end{equation}
The inequality for $\tau_{p}$ becomes strict when enforcing sufficient conditions for linear stability of equilibrium states in Minkowski spacetime derived in Ref.~\cite{Bemfica:2020zjp} [Eqs.~(49)-(50) therein]. 

In this work, we consider the following three choices of causal frame
\begin{subequations}
\begin{alignat}{21}
    &\text{A:}\quad&&\hat{\tau}_{p}=1.5, &&&\,\,\,\hat{\tau}_{e}=15,\,\,\,&&&&\hat{\tau}_{q}=20,\,\,\,&&&&&c^{2}_{\mathrm{max}} = 0.8696,&&&&&&\\
    &\text{B:}&&\hat{\tau}_{p}=2, &&&\,\,\,\hat{\tau}_{e}=20,\,\,\,&&&&\hat{\tau}_{q}=20,\,\,\,&&&&&c^{2}_{\mathrm{max}} = 0.8696,&&&&&&\\
    &\text{C:}&&\hat{\tau}_{p}=3, &&&\,\,\,\hat{\tau}_{e}=25,\,\,\,&&&&\hat{\tau}_{q}=25,\,\,\,&&&&&c^{2}_{\mathrm{max}} = 0.8621.&&&&&&
\end{alignat}\label{eq:CausalFrames}
\end{subequations}

Since the sound speed is a decreasing function of radius, causality is then guaranteed provided $c_{s}^{2}$ at the center of the background TOV solution is less than $c_\mathrm{max}^{2}$, which is the case for all three frames and for all the results presented in this paper.

\section{Formulation of the linearized problem}\label{sec:LinearProblem}
In this section, we describe the equations of motion governing the radial oscillation modes of viscous, spherically symmetric stars in the perturbative regime. We use the symbols $\delta$ and $\Delta$ to denote Eulerian and Lagrangian perturbations, respectively. 

We consider linear perturbations of equilibrium configurations of the form
\begin{equation}
    \begin{aligned}
        \nu &= \nu_{0}(r) + \varepsilon \delta\nu(t,r) \, , \\
        \lambda &= \lambda_{0}(r) + \varepsilon \delta\lambda(t,r) \, , \\
        \epsilon &= \epsilon_{0}(r) + \varepsilon \delta\epsilon(t,r) \, , \\
        u^{a} &= e^{-\nu/2}\Bigl[\Bigl(1-\varepsilon \frac{\delta\nu}{2}\Bigr)\partial^{a}_t + \varepsilon \delta{u}(t,r) \partial^{a}_r\Bigr] \, ,
    \end{aligned}
\end{equation}
where $\varepsilon \ll 1$ is a perturbative parameter controlling deviations from equilibrium. The background solution is characterized by the three functions $\nu_{0}(r),\lambda_{0}(r),$ and $\epsilon_{0}(r)$, for which we henceforth drop the subscripts. We describe the equations governing the background solution in Sec.~\ref{ssec:LinearProblem:BG} and the perturbed system in Sec.~\ref{ssec:LinearProblem:Pert}.

\subsection{Equilibrium Solutions}\label{ssec:LinearProblem:BG}
The equilibrium configuration of a cold, spherically symmetric star is governed by the Tolman-Oppenheimer-Volkoff (TOV) equations~\cite{Tolman_RTC} 
\begin{subequations}
    \begin{align}
        \frac{\ed m}{\ed r}&= 4 \pi  r^2 \epsilon,\\
        \frac{\ed p}{\ed r}&= \frac{\left(m+4 \pi  r^3 p\right) (p+\epsilon)}{r (2 m-r)},\\
        \frac{\ed \nu}{\ed r}&= \frac{2 m+8 \pi  r^3 p}{r (r-2 m)},
    \end{align}\label{eq:TOV}
\end{subequations}
where the mass function $m(r)$ is related to the metric function $\lambda(r)$ by $\lambda=-\ln{(1-2m/r)}$. The system \eqref{eq:TOV} is closed by a fourth equation, called the equation of state, which relates three thermodynamic variables. Throughout, we consider cold neutron stars described by a polytropic equation of state $p(\epsilon)=\kappa \epsilon^{1+1/n}$.

Integration of the TOV equations requires three boundary conditions. Regularity at the center of the star requires $m(0)=0$. Since \eqref{eq:TOV} is invariant under $\nu(r)\to\nu(r)+\mathrm{const}$, given a central density $\epsilon_c\equiv\epsilon(0)$, one can choose an arbitrary value of $\nu(0)$ and integrate out to the stellar surface $r=R_{S}$, where $p(R_S)=0$. Birkhoff's theorem dictates that the exterior $r>R_{S}$ solution is the Schwarzschild metric. Continuity between the interior and exterior solutions requires $\nu(R_{S})=\ln(1-2M_{S}/R_{S})$, where $M_{S}\equiv m(R_{S})$ is the gravitational mass of the star, uniquely fixing $\nu$. Given an equation of state, the TOV equations \eqref{eq:TOV} thus have a one-parameter family of solutions labeled by $\epsilon_{c}$.

\subsection{Radial Perturbations}\label{ssec:LinearProblem:Pert}

\subsubsection{Perfect Fluid and Eckart Frame}\label{sssec:LinearProblem:Eckart}
In the Eckart and perfect fluid cases, it is helpful to introduce the Lagrangian displacement $\xi$, where $\delta{u}=\partial_{t}\xi$. This allows one to solve the $tt$ and $tr$ Einstein equations to obtain
\begin{subequations}
\begin{align}
    \delta\lambda&=-8\pi r\xi\left(p+\epsilon\right){e^{\lambda}},\label{eq:Eckart:DeltaLam}\\
    \delta\epsilon&=-\frac{1}{r^{2}}\frac{\ed}{\ed r}\left[r^{2}\xi\left(p+\epsilon\right)\right].\label{eq:Eckart:DeltaEps}
\end{align}\label{eq:Eckart:DeltaVars}
\end{subequations}
The $rr$ Einstein equation can be solved for $\partial_{r}\delta\nu$, which, alongside \eqref{eq:Eckart:DeltaVars}, can be substituted into the energy-momentum conservation equation $\nabla_{\mu}T^{\mu r}=0$ to obtain the following master equation for the Lagrangian displacement
\begin{align}
    \left(\textsf{L}_{\rm{PF}}+\textsf{L}_{\rm{eck}}\right)\xi=0,\label{eq:Eckart:MasterTD}
\end{align}
where
\begin{subequations}
    \begin{align}
        \textsf{L}_{\rm{PF}}&=e^{\lambda-\nu} \partial_{t}^2 -c_{s}^2 \partial_{r}^{2}+\textsf{A}_{1} \partial_{r} + \textsf{A}_{2} ,\\
        \textsf{L}_{\rm{eck}}&=\left(\textsf{A}_{3} \partial_{r}^2+\textsf{A}_{4}\partial_{r} +\textsf{A}_{5}\right)\ast \partial_{t},
    \end{align}\label{eq:Eckart:MasterOps}
\end{subequations}
and the coefficients $\textsf{A}_{i}$ are functions of the background,
\begin{widetext}
    \begin{subequations}
        \begin{align}
            \textsf{A}_{1}&=\frac{1}{2 r} \left\{e^{\lambda} \left[\left(1-8 \pi  r^2 \epsilon\right) c_{s}{}^2+8 \pi  r^2 p+1\right]-4 r c_{s} c_{s}'-5 c_{s}{}^2-1\right\},\\
            \textsf{A}_{2}&=\frac{1}{2 r^2} \left\{2 r e^{\lambda} \left[\left(8 \pi  r^2 p+1\right) c_{s} c_{s}'+8 \pi  r \left(p-c_{s}{}^2 \epsilon\right)\right]-e^{2 \lambda} \left(8 \pi  r^2 p+1\right) \left[\left(1-8 \pi  r^2 \epsilon\right) c_{s}{}^2+8 \pi  r^2 p+1\right]\right.\nonumber\\
            &\left.\quad-10 r c_{s} c_{s}'+5 c_{s}{}^2+1\right\},\\
            \textsf{A}_{3}&=- \frac{e^{-\nu/2}}{3\left(p+\epsilon\right)} \left(3\zeta+4 \eta\right),\\
            \textsf{A}_{4}&=-\frac{e^{-\nu/2}}{6 r(p+\epsilon)} \left\{\left(3 \zeta+4 \eta\right)\left[e^{\lambda} \left(8 \pi  r^2 \epsilon-1\right)+3\right]+2\partial_{r}\left[r(3 \zeta+4 \eta)\right]\right\},\\
            \textsf{A}_{5}&=\frac{e^{-\nu/2}}{6 r^2 (p+\epsilon)} \left\{e^{\lambda} \left[\left(8 \pi  r^2 p+1\right) \partial_{r}\left(r\left(3 \zeta+4 \eta\right)\right)+\left(3 \zeta+4 \eta\right) \left(8 \pi  r^2 (5 p+\epsilon)+2\right)-9 \zeta \left(8 \pi  r^2 (2 p+\epsilon)+1\right)\right]\right.\nonumber\\
            &\left.\quad+  (3 \zeta+4 \eta)\left[e^{2 \lambda}\left(8 \pi  r^2 p+1\right) \left(8 \pi  r^2 \epsilon-1\right)+1\right]+\partial_{r}\left(r\left(3\zeta+4\eta\right)\right)+9\left(\zeta-2r\zeta'\right)\right\}.
        \end{align}
    \end{subequations}
\end{widetext}

In the time domain, the perfect fluid equation of motion $\textsf{L}_{\rm{PF}}\xi=0$ is a decoupled wave equation for the master variable $\xi$. Notice the presence of the viscous $\partial_{t}\partial_{r}^2\xi$ term in \eqref{eq:Eckart:MasterOps} --- viscosity in Eckart theory modifies the principal part of the inviscid master equation such that it is no longer a hyperbolic wave equation. We discuss boundary conditions enforced in the numerical integration of these equations in Sec.~\ref{ssec:Numerics:TD}.

We transform to the frequency domain by introducing a harmonic time dependence $\xi(t,r)=e^{-i\omega{t}}\xi(r)$. In the perfect fluid case, the resulting frequency-domain master equation is a self-adjoint boundary value problem on $r\in[0,R_{S}]$. That is, the frequency-domain equation can be cast in the form $\tilde{\textsf{L}}_{\rm{PF}}\xi(r)=\omega^{2}Q(r)\xi(r)$, where $\tilde{\textsf{L}}_{\rm{PF}}$ is of Sturm-Liouville type, the boundary conditions are $\xi(0)=0$ and $\Delta{p}(R_{S})=0$, and $Q=Q(r)$ is a weight function defined in terms of the background TOV solution~\cite{Chandrasekhar_1964}. From standard results of Sturm-Liouville theory, it follows that a radially perturbed perfect fluid star has a countably infinite number of real discrete oscillation modes (eigenvalues) $\omega^{2}_{0}>\omega^{2}_{1}>\ldots$ whose associated eigenvectors are orthogonal with respect to the weight function $Q(r)$. Linear stability of the star is determined by the fundamental mode: if $\omega^{2}_{0}>0$, the star is linearly stable to radial perturbations, otherwise generic perturbations grow exponentially on a timescale $\mathrm{Im}(\omega_{0})^{-1}$. 

The main focus of this paper is to characterize how viscosity (i) shifts the oscillation modes $\omega_{i}$ for stable equilibrium configurations, and (ii) affects the threshold and rate of gravitational collapse, i.e., the mode $\omega_{0}$ when $\rm{Im}(\omega_{0})>0$. We do this by solving the master equation \eqref{eq:Eckart:MasterTD} and the BDNK system described in the next section in the time domain. In the Eckart case, we can also directly compute eigenvalues and eigenvectors in the frequency domain, as we describe next.

As remarked above, viscosity in the Eckart frame breaks the hyperbolic structure of the time-domain perfect fluid master equation. Although the self-adjoint structure of the perfect fluid frequency-domain equation is consequently lost, we can still formulate the Eckart equation of motion as an eigenvalue problem for mode frequencies $\omega_{i}\in\mathbb{C}$. Defining $\Xi\equiv\partial_{t}\xi$ and introducing a harmonic time dependence for both $\xi$ and $\Xi$, the master equation \eqref{eq:Eckart:MasterTD} becomes a linear second-order ordinary differential equation
\begin{align}
    i \omega  \Xi&=e^{\nu-\lambda}\left[-c_{s}^{2} \xi ''+\textsf{A}_{1} \xi '+\textsf{A}_{2} \xi \right.\nonumber\\
    &\left.\qquad\qquad+\textsf{A}_{3} \Xi '' +\textsf{A}_{4} \Xi ' +\textsf{A}_{5} \Xi\right].\label{eq:Eckart:MasterFD}
\end{align}
Together, \eqref{eq:Eckart:MasterFD} and the equation $i\omega\xi=-\Xi$ can be solved as an eigenvalue problem for $\omega\in\mathbb{C}$ and the eigenvector $(\xi,\Xi)$. We describe in Sec.~\ref{ssec:Numerics:FD} the numerical methods employed to solve for the Eckart eigenvalues and eigenvectors.

\subsubsection{Causal Frames}\label{sssec:LinearProblem:CausalFrame}
The additional terms in the BDNK stress-energy tensor increase the order of the highest derivatives in the problem and complicate the structure of the lower-order Einstein equations. As a result, introducing the Lagrangian displacement no longer permits a direct time-integration of the $tr$ Einstein equation, which, in the Eckart case, allows one to eliminate $\delta{\epsilon}$ and $\delta{\lambda}$ as in \eqref{eq:Eckart:DeltaVars}. Instead, introducing $\xi$ only increases the highest derivative to third order. We therefore choose to cast our equations of motion directly in terms of $\delta{u}$, which leads to a system of hyperbolic wave equations for $\delta{u}$ and $\delta\epsilon$.

We eliminate $\partial_{r}\delta\nu$ from the system by solving the $rr$ Einstein equation. The $\theta\theta$ Einstein equation alongside the $t$ and $r$ energy-momentum conservation equations form a system of evolution equations for the perturbation variables $\mathbf{U}(t,r)\equiv[\delta{u},\delta\epsilon,\delta\lambda]^{\rm{T}}$, whose principal part has the structure
\begin{widetext}
\begin{align}
    \begin{pmatrix} 
     0 & 0 & \bullet \\ 
     0 & \bullet & \bullet \\ 
    \bullet & \bullet & \bullet 
    \end{pmatrix} \partial_{t}^{2}\mathbf{U} 
    + 
    \begin{pmatrix} 
    \bullet & 0 & \mathbf{\textcolor{blue}{0}} \\ 
    \bullet & \bullet & 0\\ 
    \bullet & 0 & 0
    \end{pmatrix} \partial_{r}^{2}\mathbf{U} 
    + 
    \begin{pmatrix} 
     0 & \bullet & \bullet \\ 
    \bullet & \bullet & \bullet \\ 
    \bullet & \bullet & \bullet 
    \end{pmatrix} \partial_{t}\partial_{r}\mathbf{U} 
    + \mathrm{L.O.T} = 0,\label{eq:BDNK:WaveSystem}
\end{align}
\end{widetext}
where we have used a $\bullet$ to denote (complicated) functions of the background TOV solution. The system \eqref{eq:BDNK:WaveSystem} consists of two coupled wave equations for $\delta{u}$ and $\delta\epsilon$ and a third equation, which, although second-order in time, is absent a $\partial_{r}^2\lambda$ term (as indicated by the blue entry) and thus lacks the structure of a genuine wave equation. Indeed, an analysis of the characteristics of the system \eqref{eq:BDNK:WaveSystem} reveals that it has two non-propagating modes, i.e., two identically zero characteristic speeds, one of whose associated eigenvector corresponds to a perturbation in $\partial_{r}\delta\lambda$. Our numerical integration of the BDNK equations leverages the presence of these non-propagating modes by transforming to a system consisting of two wave equations for $\delta{u}$ and $\delta\epsilon$ and a third constraint equation for $\delta\lambda$. We describe this system and its numerical integration in further detail in Sec.~\ref{ssec:Numerics:TD}.

The remaining characteristic speeds of the BDNK system \eqref{eq:BDNK:WaveSystem} are given by
\begin{align}
    \lambda_{\pm}^2&=e^{\nu-\lambda}\left(\frac{\Lambda_{1}\pm\sqrt{\Lambda_{1}^2-\Lambda_{0}}}{\Lambda_{2}}\right),\label{eq:BDNK:CharSpeeds}
\end{align}where
\begin{subequations}
    \begin{align}
        \Lambda_{0}&=4 c_{s}^{2} \left(p+\epsilon\right)\tau_{e} {\tau_{q}}^2  \left[\left(p+\epsilon\right){\tau_{p}}-\zeta-\frac{4}{3}\eta\right],\\
        \Lambda_{1}&=\tau_{e}\left[\left(p+\epsilon\right)c_{s}^2\tau_{q}+\zeta+\frac{4}{3}\eta\right]+\left(p+\epsilon\right)\tau_{p}\tau_{q},\\
        \Lambda_{2}&=2 \left(p+\epsilon\right){\tau_{e}} {\tau_{q}}.
    \end{align}
\end{subequations}
Perturbations propagate causally if $\lambda^{2}_{\pm}<e^{\nu-\lambda}$. Sufficient conditions for this criterion are
\begin{subequations}
    \begin{gather}
        \tau_{q},\tau_{p},\tau_{e} > 0 \, , \quad \eta,\zeta>0,\\
        \Lambda_{1}^{2}\geq\Lambda_{0}\geq0,\label{eq:BDNK:Causality1}\\
        2\Lambda_{1}\leq\Lambda_{2}.\label{eq:BDNK:Causality2}
    \end{gather}\label{eq:BDNK:Causality}
\end{subequations}
Substituting the class of frames \eqref{eq:BDNK:Frame} into \eqref{eq:BDNK:Causality} yields the causality constraints \eqref{eq:Causality}. Constraint \eqref{eq:BDNK:Causality1} is equivalent to the constraint given in Eq.~21b in Ref.~\cite{Bemfica:2020zjp} while \eqref{eq:BDNK:Causality2} is similar in structure to Eq.~21c therein. Note the general BDNK causality constraint $\rho\tau_{q}>\eta$ is absent here. This was instead recovered for even parity non-radial perturbations in Ref.~\cite{Redondo-Yuste:2024vdb}, indicating its absence stems only from the fact that we restrict the allowed class of perturbations to the system. 

\section{Numerical Methods}\label{sec:Numerics}
In this section, we describe the numerical methods we employ to solve the equations of motion governing radial perturbations of viscous stars discussed in Sec.~\ref{sec:LinearProblem}. We detail our frequency-domain methods, implemented only for Eckart stars, in Sec.~\ref{ssec:Numerics:FD} and our time-domain methods for both Eckart and BDNK stars in Sec.~\ref{ssec:Numerics:TD}. 

\subsection{Frequency Domain}\label{ssec:Numerics:FD}
We compute the radial oscillation modes of Eckart stars in the frequency domain by solving the system of equations given by $i\omega\xi=-\Xi$ and \eqref{eq:Eckart:MasterFD} as an eigenvalue problem for the eigenvalue $\omega\in\mathbb{C}$ and eigenvector $(\xi,\Xi)$. Our numerical scheme combines two distinct approaches: a matrix method and a shooting method.

The matrix method writes the system of equations in the form $\mathbf{A}\vec{x}=\lambda \vec{x}$ and numerically computes the eigenvalues and eigenvectors of the matrix $\mathbf{A}$. The matrix $\mathbf{A}$ is constructed as follows. We first compute the background TOV solution on a dense grid and interpolate $(p,\epsilon,\lambda,\nu)$ using the spline-interpolation routines provided by the package \texttt{Dierckx.jl}. We then discretize the system of equations on a uniform $N$-point grid spanning $r\in[0,R_{S}]$. Regularity at the center of the star requires $\xi(0)=0$ and $\Xi(0)=0$, so we seek the solution at points $j\in\{2,\ldots,N\}$. Equation \eqref{eq:Eckart:MasterFD} is discretized on this grid using centered differences. In combination with $i\omega \xi=-\Xi$, this gives a total of $2(N-1)$ linear discretized equations in the variables $\xi_{i}$ and $\Xi_{i}$, from which one can read off the matrix $\mathbf{A}$. This system is an eigenvalue problem for the eigenvectors $x=(\xi_{j},\Xi_{j})$ and eigenvalues $\lambda=i\omega$. The discretized equations at $j=N$ also use centered differences, meaning that one couples the system to the off-grid solutions $(\xi_{N+1},\Xi_{N+1})$. We have found our implementation to be most stable when eliminating these off-grid solutions by directly enforcing the boundary condition $\Delta{p}=0$ at the stellar surface. We do this by expanding $\Delta{p}=\delta{p}+\xi p'+\mathcal{O}(\xi^2)$, replacing $\delta{p}=c_{s}^{2}\delta\epsilon$ and $\delta{\epsilon}$ with \eqref{eq:Eckart:DeltaEps}, discretizing using centered differences and solving for $\xi_{N+1}$, which can then be eliminated from the system of equations. Taking the time derivative allows one to eliminate $\Xi_{N+1}$ in exactly the same way. With the matrix $\mathbf{A}$ so constructed, we then compute its eigenvalues and eigenvectors using the package \texttt{LinearAlgebra.jl}.

The matrix method provides a first approximation to the actual modes of the star. We refine our results by using these values as an initial guess for a shooting method. This method consists of directly solving Eq.~\eqref{eq:Eckart:MasterFD} as an ordinary differential equation (ODE) in $\xi$ (replacing $\Xi=-i\omega \xi$), using as initial conditions $\{\xi(0),\xi'(0)\} = \{0, 1\}$, and integrating up to the surface of the star. If the given $\omega$ is an eigenvalue of the differential operator, then the discretized solution should satisfy $\Delta{p}(R_{S})=0$. This constitutes a two-dimensional nonlinear root-finding problem. 

Our algorithm proceeds as follows. First, we numerically solve the TOV equations using an explicit fourth-order Runge-Kutta (RK$4$) integration scheme with a highly accurate step size of $h=1\,\mathrm{cm}$ and terminate the integration when $p=p(0)\times 10^{-8}$, which defines the stellar radius $r=R_{S}$. Next, we construct the matrix $\mathbf{A}$ and compute its eigenvalues using a grid of $N=500$ points. The resulting eigenvalues are used as an initial guess for the shooting method. We integrate the master equation using an explicit RK$4$ method and use the nonlinear root-finder provided by \texttt{NonlinearSolve.jl}. For all the eigenvalues computed in the frequency domain that we analyze in this paper, we have checked that the frequencies we obtain converge in the sense that $|\tilde{\omega}_{2}-\tilde{\omega}_{1}|<|\tilde{\omega}_{1}-\tilde{\omega}_{0}|$, separately for $\tilde{\omega}=\mathrm{Re}(\omega)$ and $\tilde{\omega}=\rm{Im}(\omega)$, where $\omega_{n}$ is a frequency computed with step size $h_{\mathrm{FD}}/2^{n}$. In all cases, $h_{\rm{FD}}\leq 5\,\mathrm{m}$.

\subsection{Time Domain}\label{ssec:Numerics:TD}
We discretize the equations of motion in the time domain on a uniform computational grid with radial step size $h$ and time step $\Delta t = 0.05 h$ (ensuring the Courant-Friedrichs-Lewy condition is satisfied). We discretize spatial derivatives with second-order centered finite differences, except at the stellar surface where we use second-order backwards differences. We describe in each subsection our treatment at the center of the star. For the time-domain simulations, we take $T=20\,\mathrm{ms}$ and perform convergence tests for three grid spacings $h\in\{5,10,20\}\,\mathrm{m}$. The background TOV equations are integrated with a step size of $h=0.1\,\mathrm{m}$ up until $r=R_{S}$, where $p=p(0)\times 10^{-6}$.

\subsubsection{Eckart Frame}
Regularity of solutions to the master equation \eqref{eq:Eckart:MasterTD} at the center of the star requires $\xi(0)=0$ and $\dot{\xi}(t,0)=0$. We integrate the master equation forward in time from initial data $\xi(0,r)$ and $\dot{\xi}(0,r)$ using an RK$4$ time integrator. In all our simulations, we choose our initial data to be stationary, i.e., $\dot{\xi}(0,r)=0$.

\subsubsection{Causal Frames}\label{sssec:Numerics:TD:BDNK}
We have found direct numerical integration of the system \eqref{eq:BDNK:WaveSystem} to be prone to numerical instabilities. In particular, we attempted to numerically integrate \eqref{eq:BDNK:WaveSystem} in time using an explicit RK$4$ scheme. We found that the resulting discretized solutions generally blow up at the boundaries of the spatial domain much faster than the dynamical timescales of the system. Following \cite{Bernuzzi:2008fu, Nagar:2004ns}, we utilize the fact that \eqref{eq:BDNK:WaveSystem} has non-propagating modes and transform to a system in which $\delta\lambda$ is evolved as a constrained degree of freedom and $(\delta{u},\delta\epsilon)$ by wave equations. The particular forms of these partial differential equations are lengthy so we relegate them, alongside a more detailed description of how they are obtained, to Appendix~\ref{app:BDNK}. Importantly, we have found the numerical time-integration of this system of equations to be more numerically stable, and, as described in more detail in Appendix~\ref{app:Convergence}, we have checked that the discretized solutions we obtain also solve the original system \eqref{eq:BDNK:WaveSystem}.

We solve a coupled system of equations governing a BDNK star consisting of a first-order constraint equation and two second-order wave equations. To numerically integrate these equations in time, we must specify initial data for the variables $(\delta\lambda,\delta\epsilon,\partial_{t}\delta{\epsilon},\delta{u},\partial_{t}\delta{u})$. We choose to specify initial data for $\delta{\epsilon}$,  $\delta{u}$ and their time derivatives. At $t=0$, the first-order equation becomes an ODE in $\delta{\lambda}$ which we solve numerically for $\delta\lambda(0,r)$ using an explicit RK$4$ integration starting at $r=0$, where regularity requires $\delta\lambda(t,0)=0$. Throughout, we assume stationary initial data, i.e., $\partial_{t}{\delta\epsilon}(0,r)=0$ and $\partial_{t}{\delta{u}}(0,r)=0$.

The constrained system is evolved in time using an implicit Crank-Nicholson scheme. At the surface of the star, we use second-order backwards differences. Regularity of solutions at the stellar center requires $\delta{u}(t,0)=0$, so we need only solve for $\delta{\epsilon}$ at $r=0$. The continuum equation at the center of the star that we solve for $\delta\epsilon(t,0)$ is obtained by expanding all the variables in the form $v(t,r)=v_{0}(t)+v_{1}(t)r+v_{2}(t)r^{2}+\ldots$. Regularity conditions for each variable are enforced by, for example, setting $v_{0}=0$ for $\delta{u}$. We evaluate the resulting equations at $r=0$ and take the equation featuring $\partial_{t}^{2}\delta{\epsilon}(t,0)$. This equation is discretized using second-order forward differences. We solve the resulting linear system of equations for the discretized variables at the advanced time step using \texttt{LinearAlgebra.jl}. We include Kreiss--Oliger (KO) dissipation~\cite{Kreiss_1973} with coefficient $0.2$ to suppress high-frequency components in the solution which otherwise cause the numerical simulation to crash at early times.

\section{Results}\label{sec:Results}
In this section, we explore the radial modes of viscous stars in Eckart and BDNK hydrodynamics. We consider two polytropic equations of state (EoS) $p=\kappa \epsilon^{1+1/n}$ with
\begin{subequations}
    \begin{alignat}{6}
        &n=1,\quad &&\kappa=100\,\mathrm{km}^2\quad &&&(\mathrm{EoS}\,\mathrm{A}),\\
        &n=0.8,\quad &&\kappa=700\,\mathrm{km}^{2.5}\quad &&&(\mathrm{EoS}\,\mathrm{B}),
    \end{alignat}\label{eq:EoS}
\end{subequations}

Computations in nuclear physics indicate that bulk viscosity is the leading viscous contribution for out-of-equilibrium neutron stars, especially in the presence of hyperons and strange quarks, due to the importance of weak-interaction driven nuclear processes such as modified Urca reactions~\cite{Ghosh:2025glz, Ghosh:2025wfx, Alford:2018lhf, Alford:2020lla, Most:2021zvc, Most:2022yhe}. For this reason, and to simplify our discussion, we set $\eta=\zeta/10$ throughout so that bulk viscosity is the leading dissipative contribution. Numerical simulations tracking these modified Urca reactions during neutron star mergers predict bulk viscosity to be as large as $\zeta \sim 10^{28}\viscgs$ during the late inspiral, and up to $\zeta \lesssim 10^{30}\viscgs$ during the merger. To aid comparison to these values, we often parametrize our results in terms of the dimensionful value of the bulk viscosity at the center of the star, $\zeta_c = \hat{\zeta}\epsilon_c (1+\kappa \epsilon_c^{1/n})L c_s^2$.

We highlight that the radial modes of a viscous star are \emph{complex}, unlike for a perfect fluid, due to the presence of dissipation. We split these frequencies into a (real) oscillation frequency $f$, and a damping time $\tau$, where 
\begin{equation}
    \omega = 2\pi f - i/\tau \, .
\end{equation}

\begin{figure*}[hbt!]
    \centering
    \includegraphics[width=1.0\textwidth]{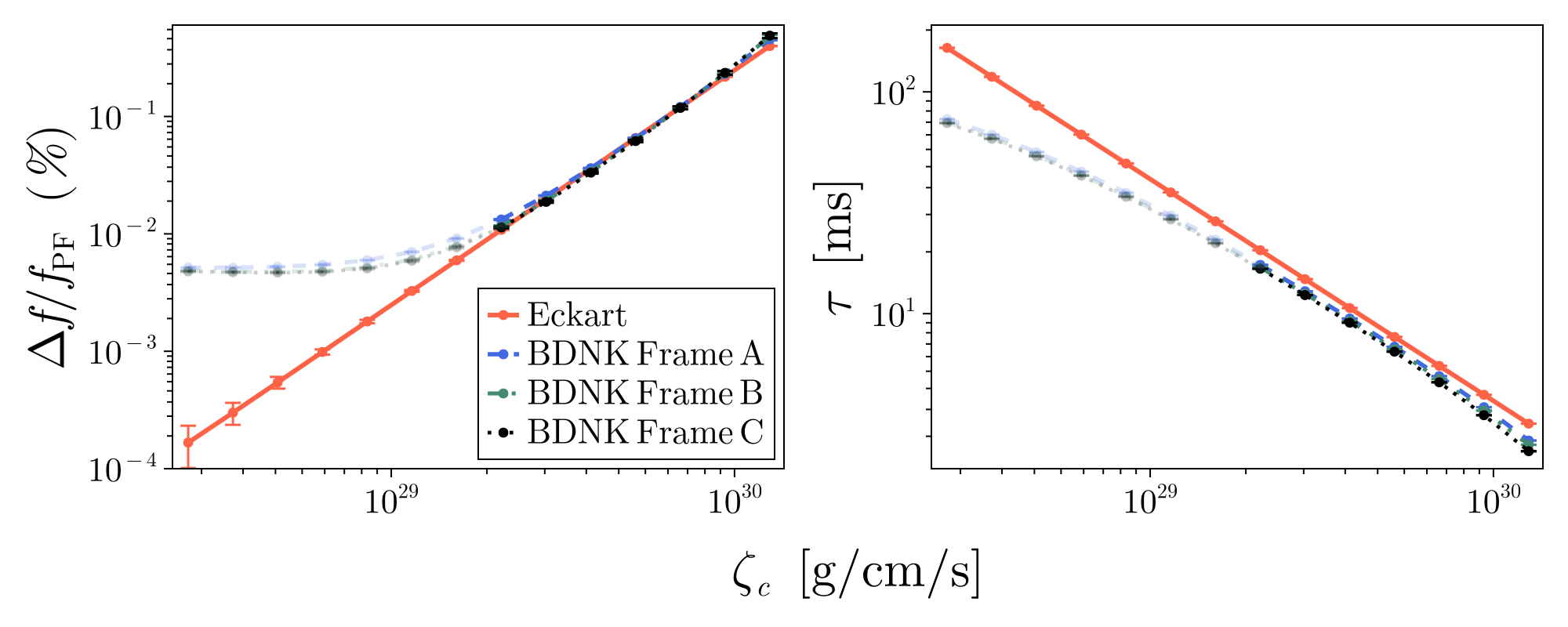}
    \caption{The effect of viscosity on the fundamental mode of neutron stars with central bulk viscosities $\zeta_{c}\lesssim10^{30}\viscgs$. \textbf{Left:} fractional frequency shifts $\Delta{f}/f_{\rm{PF}}$, where $\Delta{f}=f_{\mathrm{PF}}-f_{\mathrm{visc}}$, as a function of the central bulk viscosity. \textbf{Right:} corresponding damping times in milliseconds. The red curves are extracted from damped sinusoid fits to time-domain Eckart simulations. The blue, green and black curves are extracted from fits to time-domain BDNK simulations in the three causal frames A, B, and C, respectively, listed in \eqref{eq:CausalFrames}. For $\zeta_{c}\sim10^{30}\viscgs$, viscosity shifts the fundamental mode by $\sim1\%$. The effect of changing between causal frames is much smaller than the effect of including viscosity, indicating that these stars remain in the regime of validity of BDNK theory. There is also excellent agreement between Eckart and BDNK except in the limit $\hat{\zeta}\to0$, where numerical viscosity dominates the physical viscosity in our BDNK simulations, causing the (faint) plateau of the BDNK curves.}
    \label{fig:frequency_shifts_zeta}
\end{figure*}

Unless stated otherwise, we consider stars with EoS A, central density $\epsilon_{c} = 5.5\times 10^{15}\mathrm{g/cm^3}$ (which lies on the stable branch), and the causal frame A listed in~\eqref{eq:CausalFrames}. However, our results, as illustrated in Fig.~\ref{fig:frequency_shifts_zeta}, exhibit frame robustness, i.e., changing between causal hydrodynamic frames does not significantly modify the results. 

The remainder of this section is structured as follows. In Sec.~\ref{ssec:CodeValidation}, we verify agreement between our time- and frequency-domain methods. We recover results previously reported in the literature in the perfect fluid limit, validating the methods implemented in this paper. In Sec.~\ref{ssec:FreqShifts}, we investigate how radial oscillation modes are shifted as a function of the star's viscosity and compactness. We consider the effect of viscosity near and beyond the threshold of gravitational collapse in Sec.~\ref{ssec:UnstableBranch}, and conclude with Sec.~\ref{ssec:NonHermitian}, in which we study qualitative signatures of non-hermitian dynamics. Convergence tests for some of our simulations are presented in Appendix~\ref{app:Convergence}.

\subsection{Method Comparison}\label{ssec:CodeValidation}

\begin{figure}[hbt!]
    \centering
    \includegraphics[width=\columnwidth]{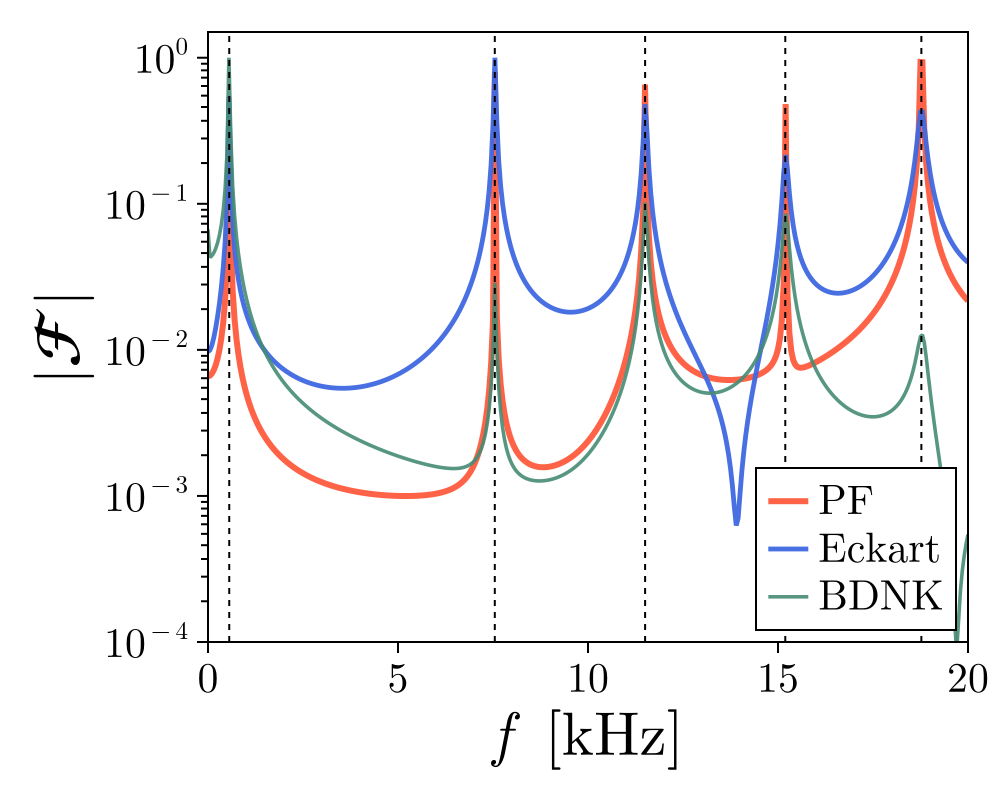}
    \caption{Mode content of Gaussian initial data evolved by the perfect fluid (red), Eckart (blue) and BDNK (green) equations  with a ``small'' central bulk viscosity, $\zeta_{c} = 1.9\times10^{29}\viscgs$. The vertical dashed lines represent the frequencies of the first five radial modes computed using the Eckart frequency-domain shooting method. We see excellent agreement between the time-domain and frequency-domain results. Time domain convergence is demonstrated in Fig.~\ref{fig:convergence}.}
    \label{fig:gaussian_td_fd_comparison}
\end{figure}

\begin{figure}[hbt!]
    \centering
    \includegraphics[width=\columnwidth]{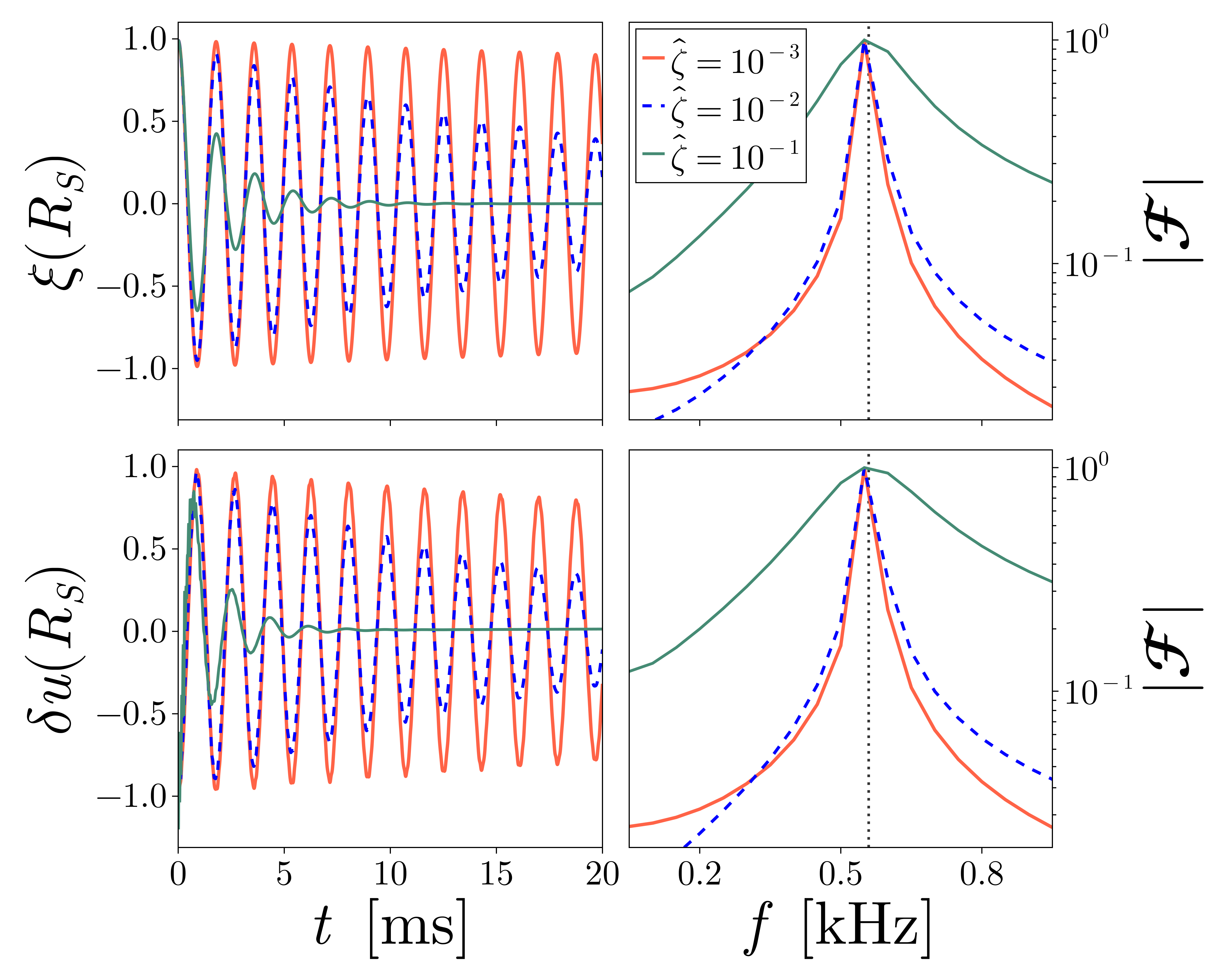}
    \caption{Lagrangian displacement (upper left) and velocity perturbation (lower left) in Eckart and BDNK simulations of fundamental mode initial data for central bulk viscosities $\zeta_{c} = 1.9\,\hat{\zeta}\times10^{31}\viscgs$ (see legend). Fourier transforms of the time series are shown in the right column, wherein the vertical line is the fundamental mode computed using the Eckart frequency-domain shooting method with $\hat{\zeta}=10^{-3}$. Significant damping occurs for the two largest viscosities, while the viscosity-induced shifts in $f$ are on the sub-Hz level.}
    \label{fig:td_fd_comparison}
\end{figure}

We begin by demonstrating agreement between our frequency- and time-domain implementations. We consider the time-domain evolution of a stationary Gaussian pulse. This perturbation excites several radial oscillation modes whose frequencies we extract by taking a Fourier transform of a sufficiently long evolution. These frequencies are compared and shown to agree exquisitely with our frequency-domain implementation. Moreover, our results agree to within the sub-percent level with the values reported in Ref.~\cite{Kokkotas:2000up} in the perfect fluid limit.

Our initial data is constructed from Gaussian pulses of the form 
\begin{align}
    \mathcal{G}(r;A,r_{0},w)&=A\exp\left(-\frac{(r-r_{0})^2}{w^{2}}\right),
\end{align}
with $(A,r_0,w)=(0.1,4,0.5)\,\mathrm{km}$. For the Eckart simulations, we consider stationary initial data with $\xi(0,r) = \mathcal{G}(r)$. For the BDNK simulations, we consider $\delta{u}(0,r)=\mathcal{G}(r)$ and $\delta\epsilon(0,r)=\mathcal{G}(r)$\footnote{We introduce additional dimensionful factors to have the correct units, e.g., $\delta\epsilon \sim \mathrm{km}^{-2}$.}.

Figure~\ref{fig:gaussian_td_fd_comparison} demonstrates the excellent agreement between the oscillation frequencies $f$ in the time-domain evolution and those computed using the frequency-domain code. The figure shows the normalized Fourier transform of $\xi(t,R_{S})$ (for the perfect fluid and Eckart stars) and $\delta{u}(t,R_{S})$ (for the BDNK stars) as solid lines. The dashed vertical lines correspond to the Eckart modes computed using the frequency-domain code. Implicit in this figure is that the shifts in the oscillation frequency $f$ are small in percentage terms for this value of the bulk viscosity --- we explore the dependence of these shifts $\Delta{f}$ on the viscosity and compactness in the next section. 

For the sake of reproducibility, we include in Appendix~\ref{app:Tables} tabulated values of the radial modes we obtain for different frames, stellar compactness, viscosities and Kreiss--Oliger coefficient values. In particular, frequencies computed in the time and frequency domains are compared in Table~\ref{tbl:TDFreqs}.

\subsection{Viscosity Shifts and Damps Radial Modes}\label{ssec:FreqShifts}

\begin{figure*}[hbt!]
    \centering
    \includegraphics[width=\textwidth]{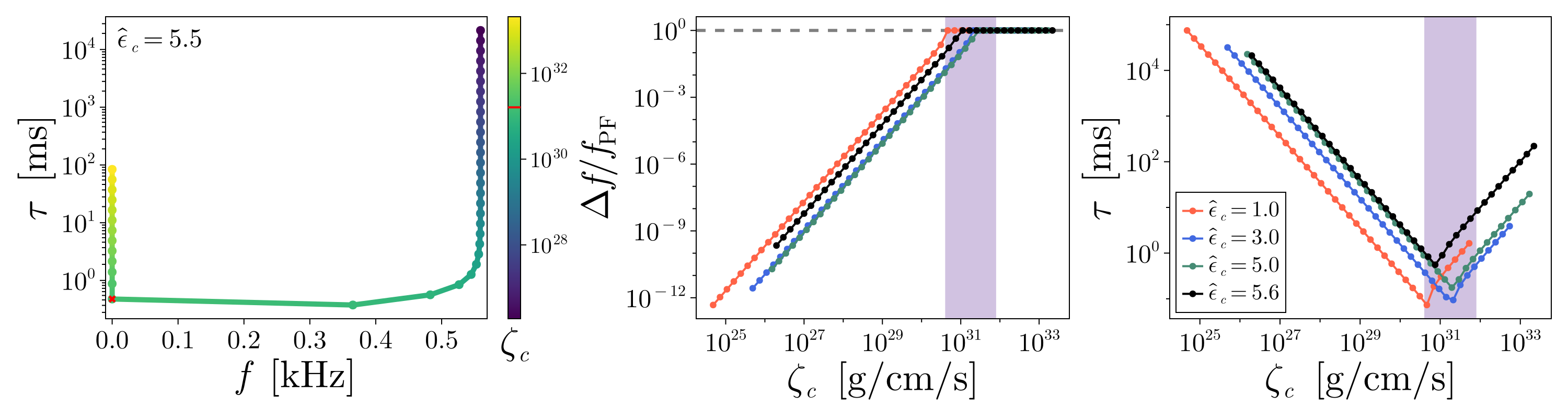}
    \caption{Transition to an overdamped fundamental mode in the large-viscosity regime of Eckart stars with central densities $\epsilon_{c}=\hat{\epsilon}_{c}\times10^{15}\epscgs$. \textbf{Left:} fundamental mode eigenvalue $\omega=2\pi{f}-i/\tau$ in the complex plane as a function of the central bulk viscosity $\zeta_{c}$ with $\hat{\epsilon}_{c}=5.5$. As $\hat{\zeta}\to\infty$, the mode becomes aperiodically damped on infinitely long time scales. Similar behavior
    occurs for the higher overtones (see Tables~\ref{tbl:FDFreqsA}--\ref{tbl:FDFreqsB}). \textbf{Middle:} fractional shifts $\Delta{f}/f_{\rm{PF}}$, where $\Delta{f}=f_{\rm{PF}}-f_{\rm{eck}}$, as a function of $\zeta_{c}$ for several central densities (see the legend in the right panel). The dashed horizontal line corresponds to unity, i.e., where the modes lie on the imaginary axis in the complex plane. \textbf{Right:} corresponding damping times for the modes shown in the middle panel. The transition to overdamped modes happens for all considered stellar models at central bulk viscosities within roughly an order of magnitude of $\zeta_{c}\sim5\times10^{31}\viscgs$, shaded in purple in the figure. For EoS B \eqref{eq:EoS}, a transition to overdamped modes also occurs around this same range of viscosities. This transition $f\to0$ with increasing $\zeta_{c}$ suggests that sufficiently viscous stars can have arbitrarily low frequency oscillation modes.}
    \label{fig:overdamping}
\end{figure*}

In this section, we study how viscosity parametrically shifts the radial oscillation modes of neutron stars. For the most compact stars studied, we find the maximum shift relative to an inviscid star to be on the percent-level when the viscosity is comparable to the post-merger bulk viscosity $\zeta_c\sim 10^{30}\viscgs$. Possibly the most impactful effect of viscosity is in damping the radial oscillation modes: for the viscosities considered here, these modes can have a damping time of the order of (tens of) milliseconds. This is consistent to within a factor of $\sim10$ with the Newtonian approximation of the $\ell=2$ non-radial mode derived in (Eq.~$21$ of) Ref.~\cite{cutler1987effect}. We also find that the star's compactness enhances the impact of viscosity. This is expected: more compact stars can also support larger gradients in, e.g., $\delta u$, resulting in larger viscous stresses even for fixed values of the transport coefficients.

We consider the time-domain evolution of single-mode initial data. This initial data is constructed by first computing the $n$-th eigenvector $\xi_n(r)$ using the Eckart frequency-domain shooting method and setting
\begin{equation}
    \begin{aligned}
        \text{Eckart:}& \quad \xi(0,r) = \mathrm{Re}[\xi_n(r)] \, , \\
        \text{BDNK:}& \quad \delta u(0,r) = \mathrm{Re}[-\xi_{n}(r)/i\omega],\, \delta \epsilon(0,r) = 0 \,.
    \end{aligned}
\end{equation}
We normalize the initial data for $\xi$ and $\delta{u}$ such that they take on a value of unity at the surface of the star. To compute $f$ and $\tau$ from the resulting time domain simulations, we fit the time series of $\xi(t,R_{S})$ and $\delta{u}(t,R_{S})$ to a superposition of damped sinusoids\footnote{We include one damped sinusoid in the perfect fluid fit and four in the Eckart and BDNK cases since single-mode initial data can significantly excite distinct modes (see Sec.~\ref{ssec:NonHermitian}). Notice that these modes have relatively large quality factors compared to, e.g., the quasinormal modes of a black hole, facilitating significant multi-mode fits.}: the error bars in the figures correspond to the standard deviation in the fitted values of $f$ and $\tau$. 

For viscosities up to $\zeta_{c}\sim10^{30}\viscgs$ the results obtained in the Eckart and BDNK stars are in excellent agreement, with their differences much smaller than the effect of including viscosity. This suggests that perturbative calculations carried out in the simpler Eckart frame may be sufficiently accurate to describe the effects of small transport coefficients. Close agreement between the results of these viscous fluid models is shown in Figs.~\ref{fig:td_fd_comparison}--\ref{fig:frequency_shifts_zeta}, which we discuss below.

The left column of Fig.~\ref{fig:td_fd_comparison} shows the evolution of fundamental mode initial data in Eckart (upper row) and BDNK (lower row) stars, with their corresponding frequency content shown in the right column. The perturbation variables $\xi$ and $\delta{u}$ undergo damped oscillations with a frequency close to that obtained from the frequency-domain (dotted line). The only noticeable difference is an enhanced damping rate for the least viscous stars, which is driven by numerical viscosity (see later discussion around Fig.~\ref{fig:frequency_shifts_zeta}).

We fit time-domain signals such as those shown in Fig.~\ref{fig:td_fd_comparison} to extract the frequency shift and damping time of the fundamental mode for a range of bulk viscosities. These results are shown in Fig.~\ref{fig:frequency_shifts_zeta} and reported in Table~\ref{tbl:TDFreqs}. Close agreement between the oscillations and damping rates in the Eckart and BDNK frames is exhibited for several viscosities. The three causal hydrodynamic frames considered are also in very close agreement, demonstrating that, over this range of viscosities, the underlying physics is not affected by a particular choice of causal frame [e.g., values of the parameters $\hat{\tau}_{i}$ in Eq.~\eqref{eq:CausalFrames}].

Over the range of viscosities considered in Fig.~\ref{fig:frequency_shifts_zeta}, viscosity can shift the fundamental radial mode by up to the percent level. In particular, for the most viscous BDNK stars with $\zeta_{c}\sim10^{30}\viscgs$, the fractional shift $\Delta{f}/f_{\rm{PF}}\approx0.5\%$. Notice that although large, such viscosities can be achieved in the aftermath of neutron star mergers. In that case, our results suggest that percent-level-accurate measurements of stellar modes may be sufficient to infer the viscous transport coefficients of nuclear matter.

\begin{figure*}[hbt!]
    \centering
    \includegraphics[width=\textwidth]{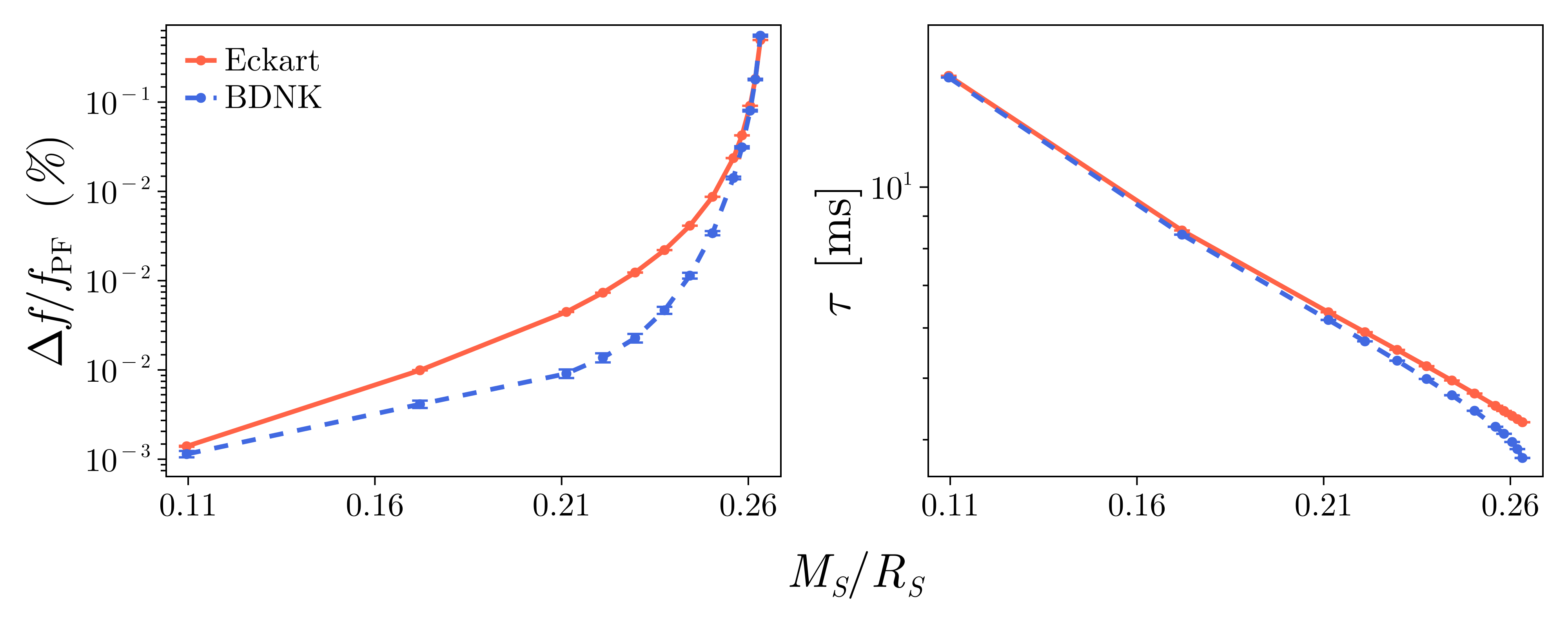}
    \caption{The effect of compactness on the fundamental radial mode of neutron stars. \textbf{Left:} fractional frequency shifts $\Delta{f}/f_{\rm{PF}}$, where $\Delta{f}=f_{\mathrm{PF}}-f_{\mathrm{visc}}$, as a function of the compactness of the star. \textbf{Right:} corresponding damping timescales for the modes shown in the left panel. The dimensionless bulk viscosity parameter is fixed to $\hat{\zeta}=0.05$. The least compact stars ($\epsilon_{c}=10^{15}\epscgs$) have $\zeta_{c}=2.4\times10^{28}\viscgs$ and the most compact stars ($\epsilon_{c}=5.5\times 10^{15}\epscgs$) have $\zeta_{c}=9.5\times10^{29}\viscgs$. The red and blue curves are obtained from damped sinusoid fits to Eckart and BDNK time-domain simulations of fundamental mode initial data. More compact stars exhibit a faster damping rate and larger fractional shifts in the oscillation frequency.}
    \label{fig:frequency_shifts_compactness}
\end{figure*}

We remark here that our time-domain implementation runs into numerical issues for small and large viscosities. For the BDNK stars, $\tau$ and $\Delta{f}$ plateau as $\hat{\zeta}\to0$, as shown by the faint part of the curves in Fig.~\ref{fig:frequency_shifts_zeta}. This is due to numerical viscosity dominating the physical viscosity (which typically happens for $\hat{\zeta}\lesssim10^{-3}$). We have verified that increasing the resolution or decreasing the KO coefficient moderates the plateau behavior, moving the with the BDNK frequency shifts closer to the Eckart values (for which we do not use KO dissipation). In this small viscosity regime, the Eckart stars have $\Delta{f}\to0$ and $\tau\to\infty$, as expected in the perfect fluid limit. Conversely, as we increase the viscosity above $\zeta_{c}\sim10^{30}\viscgs$, both our Eckart and BDNK time-domain simulations become numerically unstable. Such ``large'' viscosities coincide with a qualitative change in the behavior of the radial modes --- we investigate this large-viscosity regime in the frequency domain.

We find that Eckart stars with $\zeta_{c} \sim 10^{31}\viscgs$ are so viscous that their fundamental mode is purely imaginary, i.e., the perturbations are non-oscillatory and damped on increasingly long timescales. This effect is demonstrated in Fig.~\ref{fig:overdamping}. The existence of these aperiodic modes has a non-relativistic analogue in the oscillations of viscous bubbles, first studied by Chandrasekhar~\cite{Chandrasekhar:1959}. 

The left-most panel of Fig.~\ref{fig:overdamping} depicts the migration of the complex $n=0$ eigenvalue to the imaginary axis as the viscosity of the star is increased. This effect is shown for several central densities in the middle and right-most panels. In all cases, the fundamental mode becomes imaginary for $\zeta_{c}\sim5\times10^{31}\viscgs$ (purple shaded region). This transition to an overdamped fundamental mode appears to exhibit a weak dependence on the compactness of the star and also holds for EoS B~\eqref{eq:EoS}. These results suggest that the oscillation modes of sufficiently viscous neutron stars can be of arbitrarily low frequencies. Such low-frequency modes could contribute significantly to energy dissipation and tidal heating due to resonant mode excitation during the inspiral of such viscous compact objects. Potential astrophysical consequences of these modes, such as gravitational wave signatures in the non-radial sector, require further exploration.

Thus far, we have considered the effect of varying $\hat{\zeta}$ while keeping the central density fixed. We now consider the coupling between gravity and the fluid by instead holding $\hat{\zeta}$ fixed and varying the central density $\epsilon_{c}$, and hence the star's compactness. Note that the bulk viscosity depends on \emph{both} $\epsilon_{c}$ and $\hat{\zeta}$~\eqref{eq:CausalFrames}, so fixing $\hat{\zeta}$ does not hold fixed the bulk viscosity. 

Our results are summarized in Fig.~\ref{fig:frequency_shifts_compactness}. More compact stars exhibit (i) a larger fractional shift in the radial oscillation frequencies and (ii) more rapid damping.  As the stability threshold is approached, the Eckart and BDNK results begin to take on different growth rates, suggesting that the two fluid models may behave differently close to the threshold of gravitational collapse. This is precisely the topic of Sec.~\ref{ssec:UnstableBranch}.

\begin{figure*}
    \centering    
    \begin{minipage}[b]{0.49\textwidth}
        \centering
        \includegraphics[width=\linewidth]{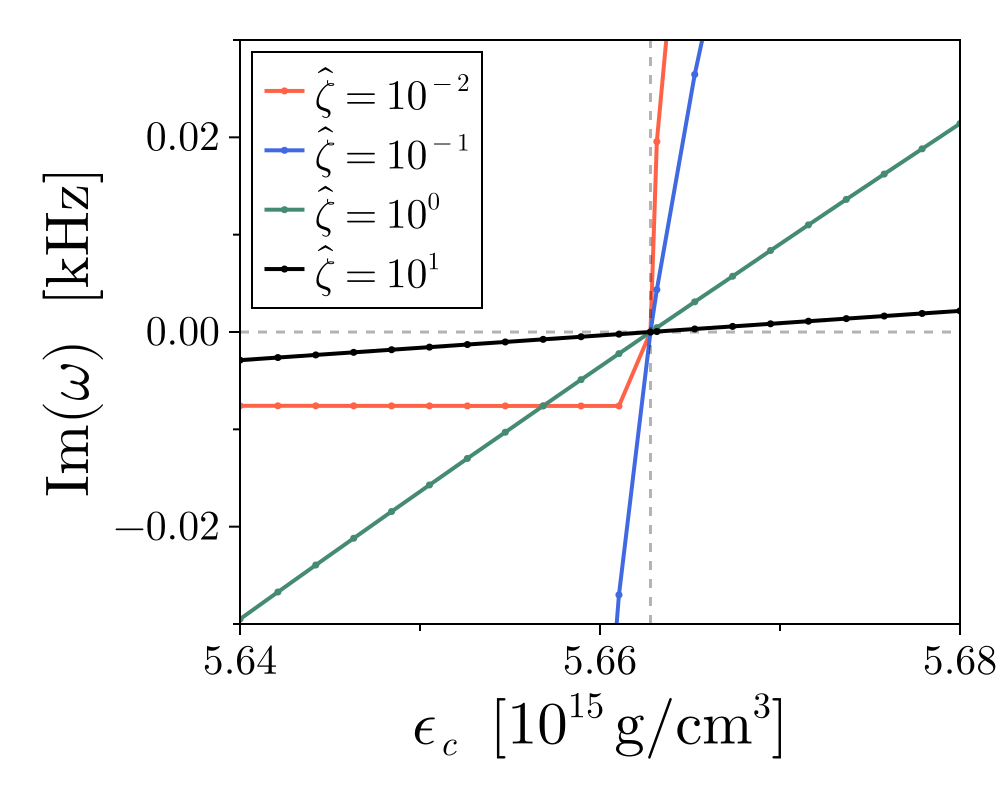} 
    \end{minipage}
    \begin{minipage}[b]{0.49\textwidth}
        \centering
        \includegraphics[width=\linewidth]{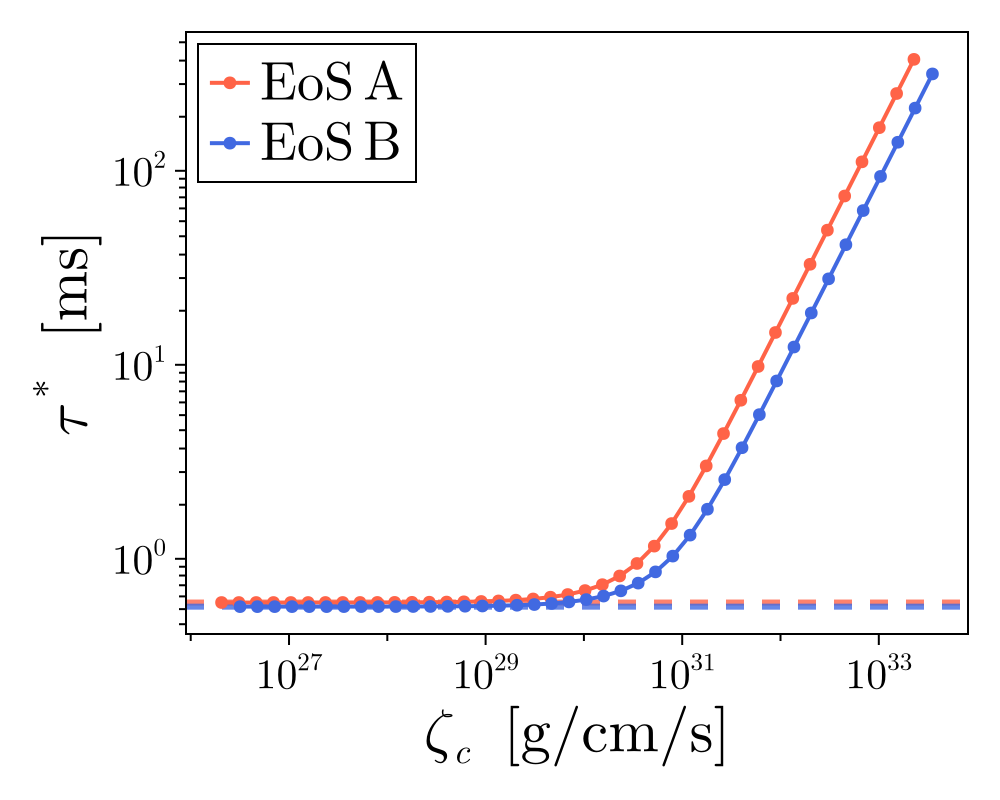}
    \end{minipage}
    \caption{The effect of viscosity on gravitational collapse in Eckart hydrodynamics. \textbf{Left:} imaginary part of the fundamental mode $\omega_0$ for central densities up to and above the threshold of collapse with fixed $\hat{\zeta}$ (see legend). The behavior of the curves differ with viscosity as they approach and emerge from the $\mathrm{Im}(\omega)=0$ axis, but they \emph{all} cross at the same central density $\epsilon_{c}^*\approx5.663\times10^{15}\epscgs$. \textbf{Right:}  instability timescale $\tau^{*}=\mathrm{Im}(\omega_0)^{-1}$ as a function of the central bulk viscosity $\zeta_{c}$. The corresponding inviscid timescales are plotted as dashed lines. The stars with EoS A and B \eqref{eq:EoS} have central densities $\epsilon_{c}=\hat{\epsilon}_{c}\times10^{15}\epscgs$, where $\hat{\epsilon}_{c}\in\{5.7,4.8\}$, respectively. While viscosity in an Eckart star does not affect the threshold of collapse, it significantly slows down the associated instability for bulk viscosity $\zeta\gtrsim10^{30}\,\viscgs$, increasing the timescale by several orders of magnitude in very viscous stars.}
    \label{fig:UnstableEckart}
\end{figure*}

Overall, we find that the effect of viscosity on the fundamental radial mode of neutron stars operates in two distinct regimes. As one ``turns on'' and increases the viscosity, the oscillation frequencies are shifted downwards and the damping timescales decrease, both scaling as a power law in $\zeta$. For bulk viscosities $\zeta \lesssim 10^{30}\viscgs$, the perfect fluid oscillation frequencies receive up to percent-level corrections and are damped on millisecond timescales. This suggests percent-level-accurate measurements of stellar oscillation modes may facilitate inference of transport properties of nuclear matter in the core of neutron stars. Further, we have found close agreement between Eckart and BDNK for small viscosities, suggesting that the simpler Eckart frame may be sufficiently accurate for calculations in this regime.

As the viscosity increases, the fundamental mode oscillation frequency becomes arbitrarily small, vanishing for $\zeta \sim 10^{31}\viscgs$. Similar behavior has already been observed in the oscillations of a non-relativistic, viscous liquid globe~\cite{Chandrasekhar:1959}. The fact that hyperonic and quark matter may probe this regime of large bulk viscosity motivates further investigation into the potential astrophysical impact of these aperiodic, overdamped modes.

\subsection{Can Viscosity Halt Gravitational Collapse? }\label{ssec:UnstableBranch}

Stars cannot be arbitrarily compact. Increasing the central density beyond a certain critical threshold results in the star being unstable: arbitrarily small radial perturbations grow exponentially in time, signaling the collapse of the star under its own gravity~\cite{Chandrasekhar_1964}. The precise threshold $\epsilon_c^*$ depends on the equation of state and has been thoroughly characterized for perfect fluid stars~\cite{Kokkotas:2000up}. This has also been studied nonlinearly~\cite{Gabler:2009yt}, showing good agreement with perturbative predictions. In this section, we study the impact of viscosity on the radial perturbations of viscous stars close to and above the threshold of gravitational collapse. 

\begin{figure}[hbt!]
    \centering
    \includegraphics[width=\columnwidth]{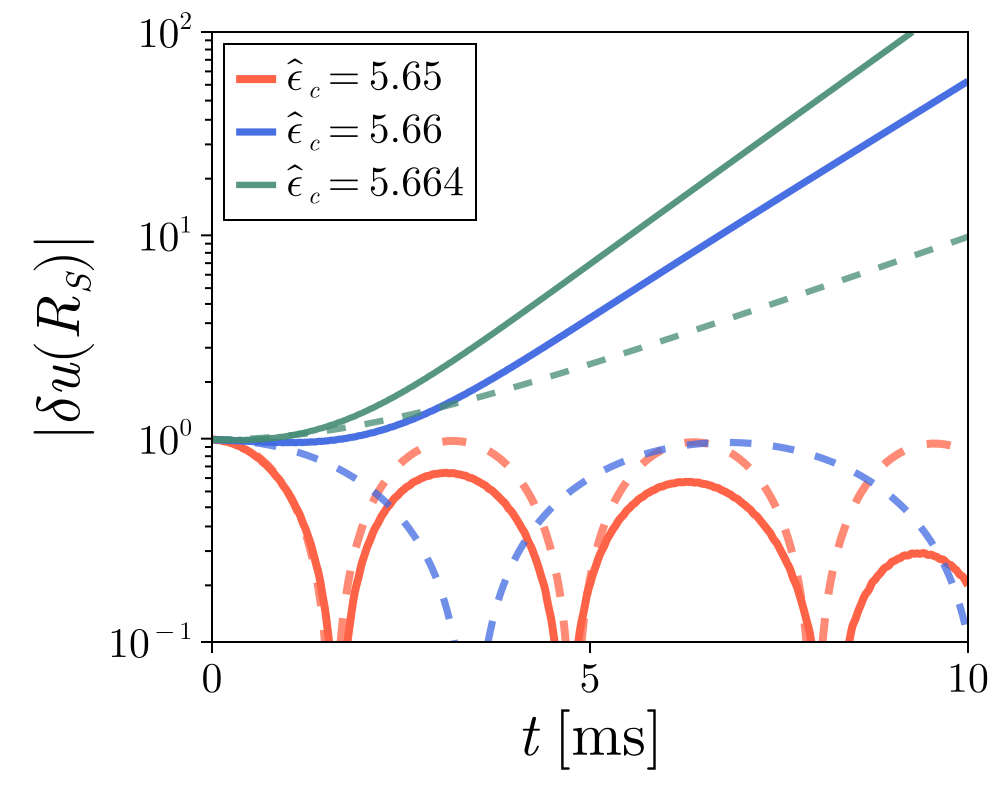}
    \caption{Time series of $|\delta{u}(t,R_{S})|$ \textbf{(BDNK, solid)} and $|\xi(t,R_{S})|$ \textbf{(Eckart, dashed)} in evolutions of fundamental mode initial data with bulk viscosity $\hat{\zeta}=10^{-3}$ and central densities $\epsilon_{c}=\hat{\epsilon}_{c}\times10^{15}\epscgs$ (see legend). The least compact viscous stars undergo damped oscillations (red curves). The BDNK star with $\hat{\epsilon}_{c}=5.66$ is unstable to radial perturbations at the linear level, while the corresponding Eckart star is stable (blue curves). Our frequency-domain code predicts the Eckart (and perfect fluid) stars are unstable above $\hat{\epsilon}_{c}\approx5.663$, which is consistent with the dashed green curve.  Viscosity in Eckart hydrodynamics does not change the linear threshold of gravitational collapse. These results suggest viscosity in BDNK theory slightly modifies this threshold.}
    \label{fig:BDNK_unstable_branch}
\end{figure}

Our key findings are as follows. First, we numerically demonstrate that viscosity in Eckart hydrodynamics does not modify the perfect fluid threshold of gravitational collapse, as has been shown through analytic arguments in Ref.~\cite{Caballero_2025}. Second, we show that viscosity can significantly affect how rapidly this instability ensues: we find that viscosity shifts the timescale of the instability by an amount which scales proportionally with $\zeta$, dominating the inviscid contribution for $\zeta\gtrsim10^{30}\,\viscgs$. We provide numerical evidence that the threshold of collapse \emph{is} modified by viscosity in BDNK hydrodynamics, but this modification is small. These results are summarized in Figs.~\ref{fig:UnstableEckart}--\ref{fig:BDNK_unstable_branch}.

The left panel of Fig.~\ref{fig:UnstableEckart} shows the imaginary part of the Eckart fundamental mode $\rm{Im}(\omega_{0})$ as a function of the central density close to the threshold of collapse. Notice that $\rm{Im}(\omega_0)<0$ corresponds to \emph{damped} modes and $\rm{Im}(\omega_0)>0$ to \emph{exponentially growing} modes. \emph{All} the curves cross the $\mathrm{Im}(\omega)=0$ axis at the same $\epsilon_{c}$, irrespective of how viscous the star is. Implementing a root-finding algorithm such that $\rm{Im}(\omega)=0$, we find that the threshold for stability is given by $\epsilon_{c}^*\approx5.663\times10^{15}\epscgs$, which agrees to within $14$ decimal places for all the viscosities (including a perfect fluid $\hat{\zeta}=0$). The onset of instability of an Eckart star with EoS B is similarly independent of the viscosity. On either side of this threshold, the gradient of the curves depend on the viscosity of the star. In particular, the instability rate $\rm{Im}(\omega_0)$ at a fixed $\epsilon_{c}>\epsilon_{c}^{*}$ decreases with bulk viscosity.

The right panel of Fig.~\ref{fig:UnstableEckart} shows instability timescale of an Eckart star as a function of the central bulk viscosity. We fix a central density for which the imaginary part is positive, i.e., the star is unstable, and study how the instability rate changes with increasing viscosity. The inviscid stars collapse on a timescale $\tau^*\sim 0.5\,\mathrm{ms}$. The viscosity-induced shift in the instability timescale scales linearly with bulk viscosity with an equation of state dependent gradient. Viscous effects dominate and slow down the collapse for bulk viscosities $\zeta_c \gtrsim 10^{30}\viscgs$. Practically, a very large bulk viscosity [$\zeta_c \gtrsim 10^{33} \viscgs$] is required in order to have an instability timescale on the order of, e.g, seconds for the stars considered here.

Lastly, we study the stability of stars close to the threshold of collapse in BDNK hydrodynamics. In Fig.~\ref{fig:BDNK_unstable_branch}, we show time-domain simulations of a low-viscosity ($\hat{\zeta}=10^{-3}$) BDNK star for three values of the central density below and above the (perfect fluid) stability threshold $\epsilon_c < \epsilon_c^*$. For comparison, we also show in dashed lines the same configurations but for Eckart stars. We find that the BDNK stars with central density $\epsilon_c = 5.66\times10^{15}\mathrm{g/cm^3} < \epsilon_c^*$ are \emph{unstable}, whereas the same configuration is stable in the Eckart and perfect fluid cases. This suggests the threshold for BDNK stars may differ from that for Eckart stars.

In summary, our results suggest viscosity is unable to prevent gravitational collapse at the linear level: we find that stars which are unstable within a perfect fluid description are also unstable in the viscous theories considered in this work. Viscosity does, however, slow down collapse: the viscous shift in the instability timescale is directly proportional to the viscosity of the star, dominating the inviscid contribution for $\zeta\gtrsim10^{30}\,\viscgs$. Our results also provide evidence that the threshold for instability may depend on viscosity in BDNK stars. This motivates a detailed follow-up study of the full nonlinear problem in order to characterize the stability of BDNK stars beyond the linear level and, importantly, to verify that the gradients of the solution do not grow too large such that the regime of validity of BDNK theory is exited. First steps in this direction were already taken in Ref.~\cite{Shum:2025jnl}.

\subsection{Viscosity and Non-Hermitian Dynamics}\label{ssec:NonHermitian}

In this section, we study qualitative signatures of non-hermitian dynamics in Eckart and BDNK time-domain simulations. We consider the evolution of fundamental mode and first-overtone initial data. For \emph{both} the Eckart and BDNK simulations, we construct initial data from Eckart eigenvectors computed in the frequency-domain to dominantly excite one of these modes. 

\begin{figure*}[hbt!]
    \centering    
    \begin{minipage}[b]{0.49\textwidth}
        \centering
        \includegraphics[width=\linewidth]{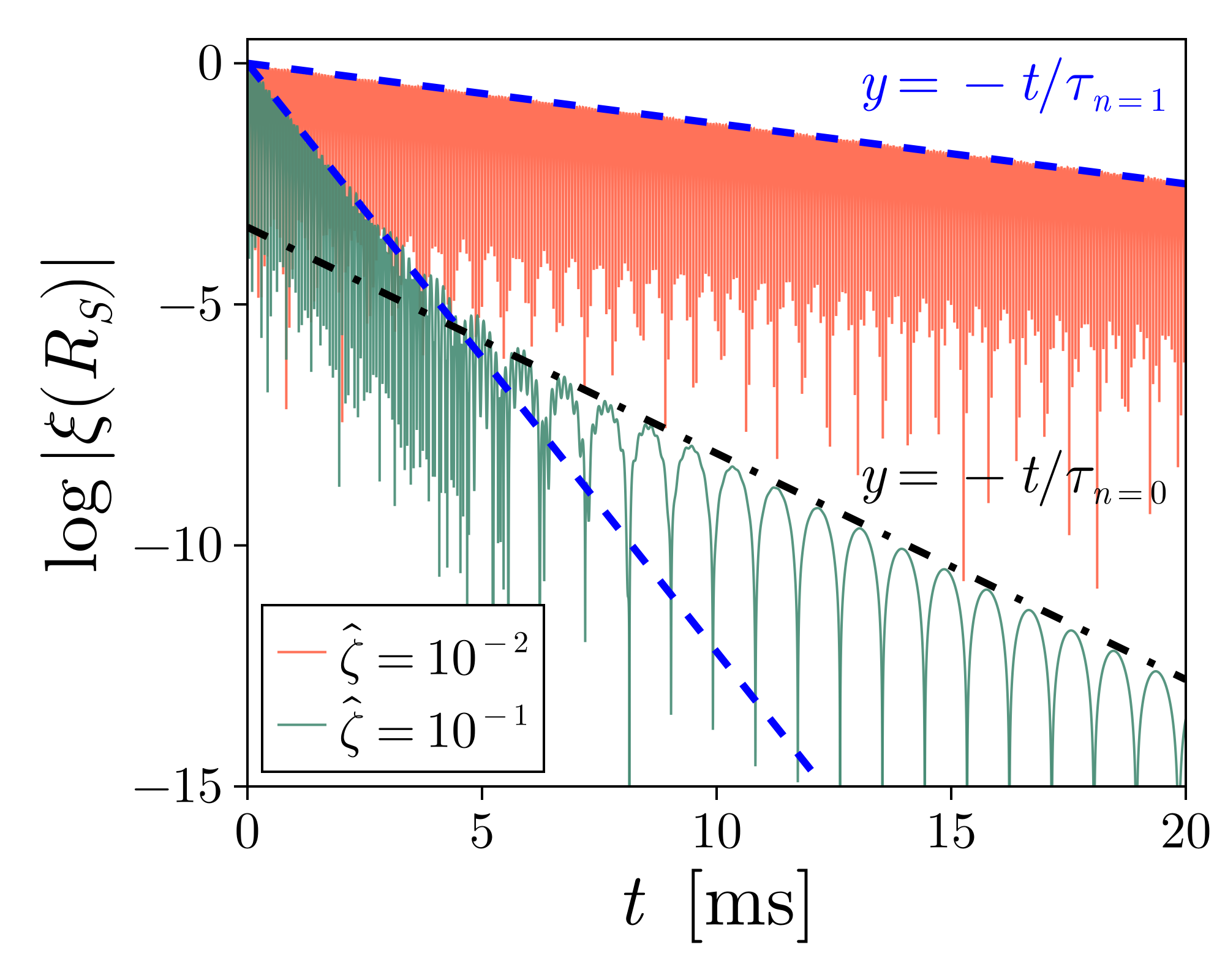} 
    \end{minipage}
    \begin{minipage}[b]{0.49\textwidth}
        \centering
        \includegraphics[width=\linewidth]{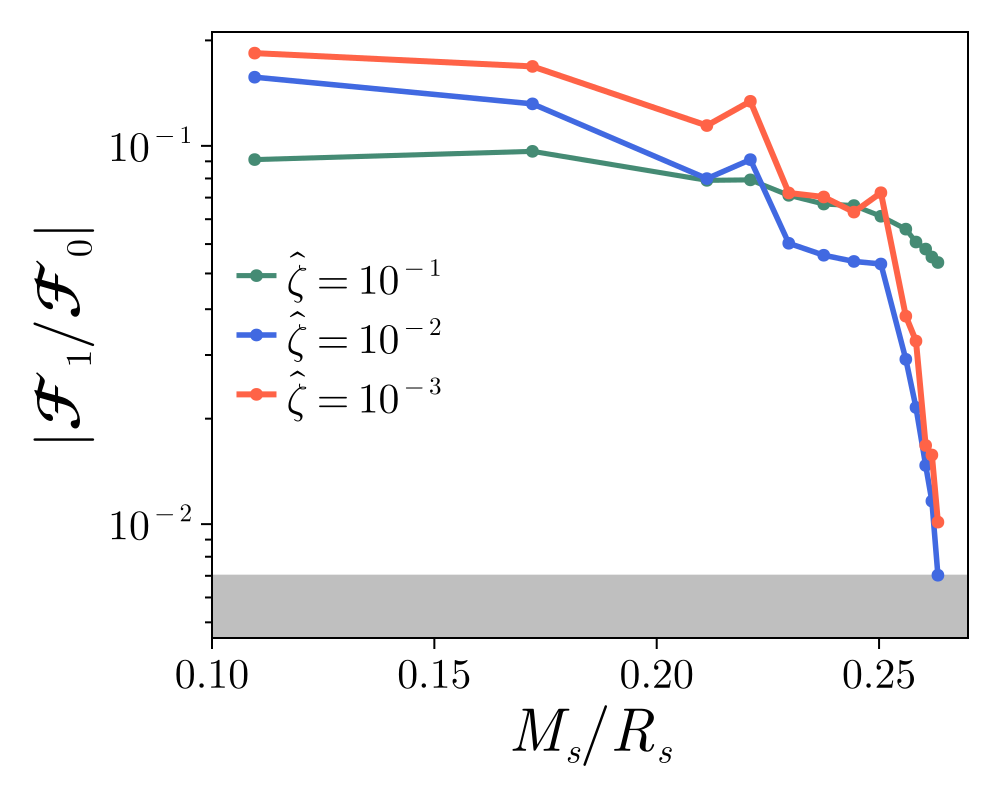}
    \end{minipage}
    \caption{Qualitative signatures of non-hermitian dynamics in Eckart and BDNK stars. \textbf{Left:} damped oscillations of $\xi(t,R_{S})$ in Eckart time-domain simulations of $n=1$ overtone initial data for two values of the central bulk viscosity $\zeta_{c} = 1.9\,\hat{\zeta}\times10^{31}\viscgs$ (see legend). The black and blue curves correspond to $y\sim\exp(-i/\tau_{n})$, where $\tau_{n}=\mathrm{Im}(\omega_{n})^{-1}$ is computed via the Eckart frequency-domain shooting method. The damped oscillations of the less viscous star are consistent with the $n=1$ eigenvalue, which only holds on short timescales $t\lesssim 5\,\mathrm{ms}$ for the more viscous star, after which the fundamental mode dominates. \textbf{Right:} amplitude of the Fourier transform of $\delta{u}(t, R_{S})$ corresponding to the $n=1$ mode relative to $n=0$ in time-domain BDNK evolutions of (Eckart) fundamental mode initial data for three bulk viscosities (see legend). Qualitatively, the structure of BDNK eigenvectors strongly changes as the stability limit is approached: the Eckart eigenvectors become better approximations in that the $n=0$ mode is more dominantly excited, becoming comparable to analogous simulations of a perfect fluid (grey shaded region). Viscous stars exhibit strong signatures of non-hermitian dynamics already at bulk viscosities $\zeta \sim 10^{30}\viscgs$.}
    \label{fig:NonHermitian}
\end{figure*}

We find that an initial perturbation of an Eckart star which dominantly excites a single overtone $n>0$ will eventually be dominated by the fundamental mode. This is not surprising in the sense that there is a hierarchy of damping rates in the mode spectrum, with the fundamental mode being the most long lived. Nevertheless, it is interesting that this occurs even if distinct modes are excited only at the level of truncation error, as is the case here since we evolve eigenvectors (which are exact up to truncation error). Our BDNK simulations suggest that the structure of the eigenvectors of the fluid depend strongly on the compactness of the star as the threshold of stability is approached. These results are shown in Fig.~\ref{fig:NonHermitian}.

For an Eckart star with $\zeta_{c}\sim10^{29}\viscgs$, evolution of $n=1$ eigenvector initial data dominantly excites the first overtone throughout the evolution, with the solution being damped on a timescale consistent with $\rm{Im}(\omega_{1})^{-1}$ computed in the frequency domain. Increasing the viscosity to $\zeta_{c}\sim10^{30}\viscgs$ results in a faster damping rate such that $n=0$ eventually dominates on the timescales of the simulation. These results are shown in the left panel of Fig.~\ref{fig:NonHermitian}. Very similar qualitative dynamics occur for corresponding simulations of a BDNK star, although one expects more significant excitation of distinct modes due to the approximate use of \emph{Eckart} eigenvectors as initial data. 

We find that $n=0$ Eckart eigenvectors more dominantly excite the fundamental mode in more compact BDNK stars. This is illustrated in the right panel of Fig.~\ref{fig:NonHermitian}, in which we plot the Fourier transform amplitude corresponding to the $n=1$ mode relative to that of $n=0$ in BDNK evolutions of (Eckart) fundamental mode initial data. Across three orders of magnitude of the bulk viscosity, the excitation of the $n=1$ mode deceases significantly as the compactness of the star approaches the stability limit, reaching the numerical level of a perfect fluid (the grey-shaded floor) at the largest compactnesses. This dependence on the compactness, however, appears weakened the more viscous the star is. We emphasize that the approximate use of Eckart eigenvectors (as well as setting $\delta{\epsilon}=0$) in the initial data prevents a more quantitative analysis of the structure of BDNK eigenvectors.

Already at viscosities $\zeta_{c}\sim10^{30}\viscgs$, Eckart and BDNK stars exhibit clear signatures of non-hermitian dynamics. For BDNK stars, the non-trivial coupling between gravity and the viscous fluid strongly affects the qualitative structure of the eigenvectors near the stability limit of the star. Studying the radial oscillations of dissipative stars can provide a useful setting for probing the impact of dissipation and non-hermiticity in dynamics of neutron stars. 

\section{Discussion}\label{sec:Discussion}

We have systematically studied the effect of viscosity on the radial oscillation modes of neutron stars. In the small to moderate viscosity regime (bulk viscosities $\zeta \lesssim 10^{30}\viscgs$), we found that the quasinormal modes obtained within Eckart hydrodynamics are in very good agreement with those obtained in the causal framework of BDNK hydrodynamics. In this regime, the real oscillation frequency is shifted by up to the percent-level, with damping times as small as $\sim10\,\mathrm{ms}$. We also found that the impact of viscosity is generally enhanced in more compact stars. These results are obtained consistently both in the time- and frequency-domains. Our code, which also reproduces with excellent accuracy known results in the inviscid limit, is publicly available in~\cite{Keeble_2026}.

At very large viscosities, the situation is quite different. For very viscous stars ($\zeta \gtrsim 10^{31}\viscgs$, seemingly rather insensitive of the compactness of the star for two polytropic EoS's), the frequency of the fundamental radial mode can be arbitrarily small, leading to a non-oscillatory overdamped evolution. This is a relativistic analogue of an overdamping effect first noticed by Chandrasekhar when characterizing the oscillations of a viscous liquid bubble~\cite{Chandrasekhar:1959}. Such low frequency radial modes could have an observational impact, such as being resonantly excited during inspirals of sufficiently viscous neutron stars. However, to assess the physical content of these modes, it is necessary to check whether they (i) are also present in causal BDNK hydrodynamics and (ii) lie in the regime of validity of the theory. Such an assessment would require probing larger viscosities in BDNK theory than our current numerical implementation can handle, in addition to tackling the nonlinear problem to robustly characterize the regime of validity of BDNK. We leave to future work a more detailed study of these low-frequency modes and their potential astrophysical consequences.

We also investigated the impact of viscosity on the collapse instability of compact neutron stars. We numerically showed that viscosity within Eckart hydrodynamics does not modify the threshold of stability, as demonstrated analytically in Ref.~\cite{Caballero_2025}. Our results provide numerical evidence that BDNK hydrodynamics (ever so slightly) modifies the exact location of the threshold. We have checked that this shift is robust against numerical dissipation, resolution, and changes of causal frame. However, further study is required to verify that this shift is physical, such as solving the BDNK equations in the frequency domain to go beyond our use of initial data constructed from Eckart eigenvectors, and exploring the fully nonlinear regime to check that gradients do not grow too large near the instability threshold.

Overall, we highlight that in the simulations presented in this work, viscosity cannot stop gravitational collapse, although it does slow down the rate of collapse. In particular, we find that the collapse timescale scales as $\tau^* \sim  \tau^*_{\rm PF} + C \zeta$, with $C$ a constant which depends on the equation of state, and the second term becomes the dominant contribution for bulk viscosities $\zeta \gtrsim 10^{30}\,\viscgs$.

Important future work would be to study nonlinear dynamics within BDNK theory, especially near the threshold of collapse. Steps in this direction in the Cowling approximation have been taken in Ref.~\cite{Shum:2025jnl}. Further, the inclusion of finite-temperature effects is also necessary to better understand the impact of viscosity on astrophysical neutron stars. Our results could be extended to include a wider variety of equations of state, more realistically modeling the properties of neutron stars compatible with the most recent observations. Finally, the most interesting extension would be to characterize the non-radial oscillation modes of neutron stars. Preliminary work in this direction has been undertaken in~\cite{Boyanov:2024jge, Bussieres:2026rnz}.

\section*{Acknowledgments}

J.~R.-Y. is indebted to Laura Michelutti and Mathilde Menu for discussions on the oscillations of a viscous bubble. 
The authors thank Yago Bea, Alejandro C\'ardenas-Avenda\~no, Vitor Cardoso, and Hengrui Zhu for valuable discussions, and Daniel Caballero, Marcelo Disconzi, Kostas Kokkotas, and Nicol\'as Yunes for comments on a draft of this manuscript.  
The Center of Gravity is a Center of Excellence funded by the Danish National Research Foundation under grant no. DNRF184. J.~R.-Y. acknowledges support by VILLUM Foundation (grant no. VIL37766).
J.~R.-Y. is supported by NSF Grants No.~AST-2307146, No.~PHY-2513337, No.~PHY-090003, and No.~PHY-20043, by NASA Grant No.~21-ATP21-0010, by John Templeton Foundation Grant No.~62840, by the Simons Foundation [MPS-SIP-00001698, E.B.], by the Simons Foundation International [SFI-MPS-BH-00012593-02], and by Italian Ministry of Foreign Affairs and International Cooperation Grant No.~PGR01167.
The Tycho supercomputer hosted at the SCIENCE HPC center at the University of Copenhagen was used for supporting this work.

\bibliography{biblio}
\clearpage

\appendix
\onecolumngrid

\section{Convergence Tests}\label{app:Convergence}

In this appendix, we present convergence tests for a subset of the time-domain simulations presented in Sec.~\ref{sec:Results}.

The uniform numerical grids in our time-domain simulations are characterized by a single discretization scale $h$. Our numerical scheme is second-order accurate, so the discretized solutions we obtain are accurate to within an $\mathcal{O}(h^{2})$ truncation error. We measure whether our discretized solutions converge to a solution of the continuum PDEs in the limit $h\to0$ by monitoring convergence properties of an ``independent residual''. That is, we verify that the discretized solution we obtain from the RK$4$ and Crank-Nicholson time-integration schemes described in Sec.~\ref{ssec:Numerics:TD} also solve the equations of motion resulting from an independent discretization, namely a leapfrog discretization in time which couples the three time levels $t_{n-1}$, $t_n$ and $t_{n+1}$. We take the solution obtained from our RK$4$ and Crank-Nicholson schemes and evaluate the leapfrog-discretized equations of motion on these solutions. We denote the resulting ``independent'' residual by $R_{h}$, which is a function of time and space on our numerical grid. We define the convergence factor
\begin{align}
    Q_{h}(t)=\frac{||R_{2h}||_{1}}{||R_{h}||_{1}},\label{eq:ConvergenceFactorN}
\end{align}where $||\cdot||_{1}$ denotes the vector one-norm. For a $p$-th order accurate numerical scheme, one expects $Q_{h}(t)\sim 2^{p}$, where $p=2$ in our case. 

\begin{figure*}[hbt!]
    \centering    
    \begin{minipage}[b]{0.49\textwidth}
        \centering
        \includegraphics[width=\linewidth]{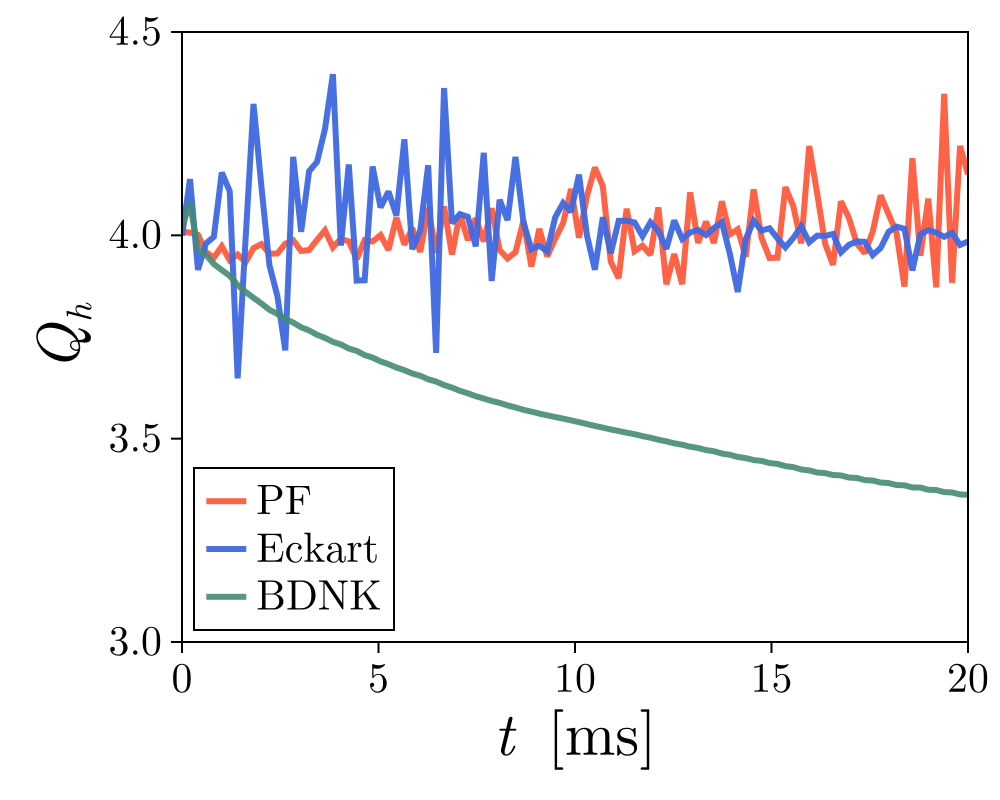} 
    \end{minipage}
    \begin{minipage}[b]{0.49\textwidth}
        \centering
        \includegraphics[width=\linewidth]{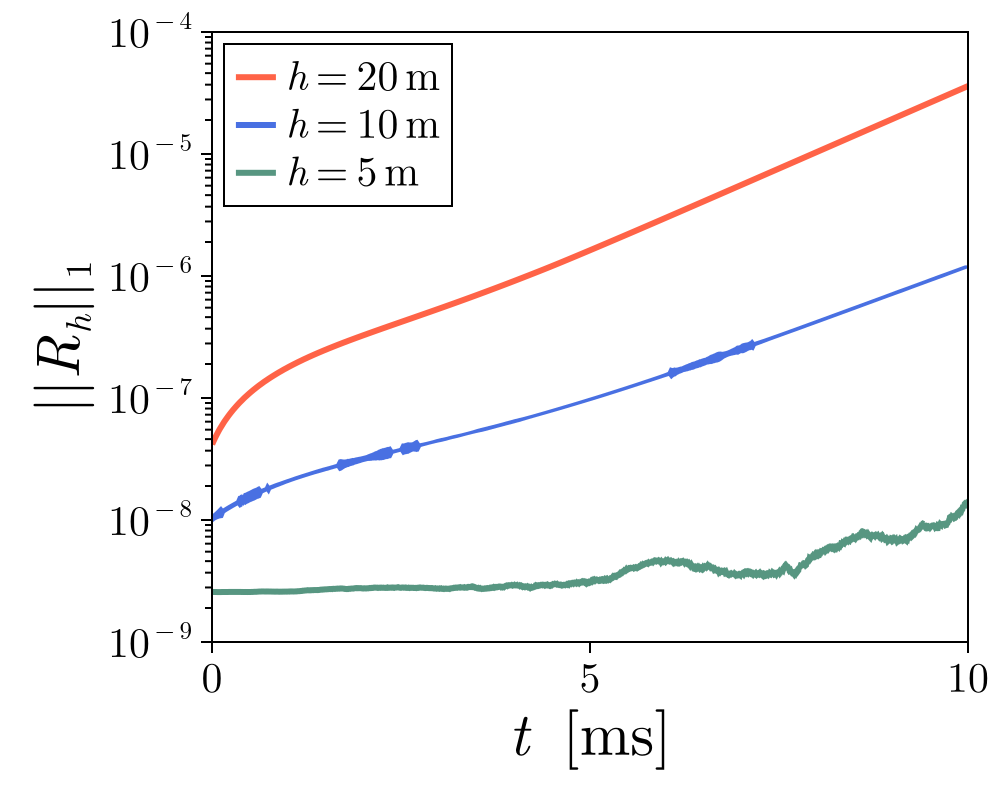}
    \end{minipage}
    \caption{Convergence of time-domain simulations. \textbf{Left:} convergence factor of the independent residual for perfect fluid (red), Eckart (blue) and BDNK (green) evolutions of Gaussian initial data shown in Fig.~\ref{fig:gaussian_td_fd_comparison}. Convergence factors $Q_{h}\sim4$ are expected for our second-order numerical schemes. The highest resolution runs have $h=5\,\mathrm{m}$. \textbf{Right:} independent residual $||R_{h}||_{1}$ in time-domain simulations of the unstable BDNK configuration shown in Fig.~\ref{fig:BDNK_unstable_branch}, which has $\hat{\zeta}=10^{-3}$ and $\epsilon_{c}=5.66\times10^{15}\epscgs$. The corresponding convergence factors have $Q_{h}>4$ at all times.}
    \label{fig:convergence}
\end{figure*}

We report convergence tests for two sets of time-domain simulations in Fig.~\ref{fig:convergence}. In the left panel we plot the convergence factor for Gaussian initial data evolved by the perfect fluid, Eckart and BDNK equations of motion as described in Sec.~\ref{ssec:CodeValidation}. We plot in the right panel convergence of the independent residual with resolution for time-domain simulations of an unstable BDNK star as described further in Sec.~\ref{ssec:UnstableBranch}. We have also verified that our numerical solutions to the constrained BDNK system also solve the original system \eqref{eq:BDNK:WaveSystem} by monitoring convergence of an independent residual of \eqref{eq:BDNK:WaveSystem} evaluated on solutions to the constrained system.

\section{Radial Oscillation Frequency Results}\label{app:Tables}
This appendix contains tabulated eigenvalues of viscous neutron star radial oscillation modes. Table~\ref{tbl:TDFreqs} contains fundamental mode eigenvalues corresponding to data points in Fig.~\ref{fig:frequency_shifts_zeta}. Table~\ref{tbl:TDFreqsKO} contains data corresponding to the BDNK eigenvalues listed in Table~\ref{tbl:TDFreqs} for three different values of the Kreiss--Oliger coefficient, showing that, for these bulk viscosities, the effect of numerical viscosity is small. Tables~\ref{tbl:FDFreqsA} and \ref{tbl:FDFreqsB} contain eigenvalues computed in the frequency domain corresponding to the first few modes of Eckart stars with two different polytropic equations of state.

\begin{table*}[hbt!]
\caption{\label{tbl:TDFreqs} Fundamental mode frequencies $f$ (kHz) and damping times $\tau$ (ms) for stars with EoS A \eqref{eq:EoS}, viscosities $\zeta_{c} = 1.9\,\hat{\zeta}\times10^{31}\viscgs$ and central density $\epsilon_{c}=5.5\times10^{15}\epscgs$. The data in the latter two sets of columns are plotted in Fig.~\ref{fig:frequency_shifts_zeta} and are obtained from fits to time-domain Eckart and BDNK simulations. To aid comparison between our numerical methods, we include in the first set of columns data computed using the Eckart frequency-domain code. We report in the first row perfect fluid frequencies computed in the time and frequency domains. The $(\cdot)$ notation denotes the first uncertain digit.}
\begin{ruledtabular}
\begin{tabular*}{\textwidth}{@{\extracolsep{\fill}}lcccccc}
& \multicolumn{2}{c}{\textrm{Eckart FD}} & \multicolumn{2}{c}{\textrm{Eckart TD}} & \multicolumn{2}{c}{\textrm{BDNK TD}} \\
\cline{2-3} \cline{4-5} \cline{6-7} 
\multicolumn{1}{c}{$\hat{\zeta}$} & \multicolumn{1}{c}{$f$} & \multicolumn{1}{c}{$\tau$} & \multicolumn{1}{c}{$f$} & \multicolumn{1}{c}{$\tau$} & \multicolumn{1}{c}{$f$} & \multicolumn{1}{c}{$\tau$} \\
\hline
$0$ & $0.559020$ & $\infty$ & $0.559253$ & $\infty$ & --- & ---\\
$0.015$ & $0.558909$ & $14.313688$ & $0.559142$ & $14.3073(5)3$ & $0.559135$ & $12.622(0)64$ \\
$0.020$ & $0.558819$ & $10.602724$ & $0.559052$ & $10.59801(2)$ & $0.55905(0)$ & $9.511(4)23$ \\
$0.027$ & $0.558653$ & $7.853858$ & $0.558886$ & $7.8503(9)2$ & $0.55888(9)$ & $7.094(3)59$ \\
$0.037$ & $0.558351$ & $5.817658$ & $0.558584$ & $5.8151(0)0$ & $0.55857(8)$ & $5.22(6)969$ \\
\end{tabular*}
\end{ruledtabular}
\end{table*}

\begin{table*}[hbt!]
\caption{\label{tbl:TDFreqsKO}BDNK fundamental mode frequency $f$ (kHz) and damping time $\tau$ (ms) data corresponding to the final set of columns in Table~\ref{tbl:TDFreqs} for different values of the Kreiss--Oliger dissipation coefficient $\alpha_{\rm{KO}}$. Varying $\alpha_{\rm{KO}}$ across these values changes the frequencies and damping times by up to $0.0016\%$ and $1.2\%$, respectively.}
\begin{ruledtabular}
\begin{tabular*}{\textwidth}{@{\extracolsep{\fill}}lcccccc}
& \multicolumn{2}{c}{$\alpha_{\rm{KO}}=0.1$} & \multicolumn{2}{c}{$\alpha_{\rm{KO}}=0.15$} & \multicolumn{2}{c}{$\alpha_{\rm{KO}}=0.2$} \\
\cline{2-3} \cline{4-5} \cline{6-7} 
\multicolumn{1}{c}{$\hat{\zeta}$} & \multicolumn{1}{c}{$f$} & \multicolumn{1}{c}{$\tau$} & \multicolumn{1}{c}{$f$} & \multicolumn{1}{c}{$\tau$} & \multicolumn{1}{c}{$f$} & \multicolumn{1}{c}{$\tau$} \\
\hline
$0.015$ & $0.559139$ & $12.472(6)76$ & $0.559137$ & $12.553(7)82$ & $0.559135$ & $12.622(0)64$ \\
$0.020$ & $0.55905(6)$ & $9.423(3)67$ & $0.55905(3)$ & $9.470(8)48$ & $0.55905(0)$ & $9.511(4)23$ \\
$0.027$ & $0.55889(6)$ & $7.04(1)417$ & $0.55889(2)$ & $7.069(8)46$ & $0.55888(9)$ & $7.094(3)59$ \\
$0.037$ & $0.55858(7)$ & $5.19(4)032$ & $0.55858(2)$ & $5.21(1)743$ & $0.55857(8)$ & $5.22(6)969$ \\
\end{tabular*}
\end{ruledtabular}
\end{table*}

\begin{table*}[hbt!]
\caption{\label{tbl:FDFreqsA} Frequencies $f$ (kHz) and damping times $\tau$ (ms) of the first few modes of Eckart stars with EoS A \eqref{eq:EoS}, viscosities $\zeta_{c} = 1.9\,\hat{\zeta}\times10^{31}\viscgs$ and central density $\epsilon_{c}=5.5\times10^{15}\epscgs$ computed in the frequency domain. The perfect fluid frequencies agree to within $\lesssim 1\%$ with Table~A.$18$ in Ref.~\cite{Kokkotas:2000up}.}
\begin{ruledtabular}
\begin{tabular*}{\textwidth}{@{\extracolsep{\fill}}lcccccc}
& \multicolumn{2}{c}{$n=0$} & \multicolumn{2}{c}{$n=1$} & \multicolumn{2}{c}{$n=2$} \\
\cline{2-3} \cline{4-5} \cline{6-7} 
\multicolumn{1}{c}{$\hat{\zeta}$} & \multicolumn{1}{c}{$f$} & \multicolumn{1}{c}{$\tau$} & \multicolumn{1}{c}{$f$} & \multicolumn{1}{c}{$\tau$} & \multicolumn{1}{c}{$f$} & \multicolumn{1}{c}{$\tau$} \\
\hline
$0$ & $0.559020$ & $\infty$ & $7.547259$ & $\infty$ &$11.502703$ & $\infty$ \\
$0.01$ & $0.558970$ & $21.318453$ & $7.547237$ & $8.193609$ & $11.502654$ & $4.317405$ \\
$0.1$ & $0.554024$ & $2.131756$ & $7.545111$ & $0.819345$ & $11.497749$ & $0.431736$ \\
$0.5$ & $0.416179$ & $0.425912$ & $7.493422$ & $0.163791$ & $11.378287$ & $0.086328$ \\
$1.0$ & $0.0$ & $0.632915$ & $0.0$ & $0.127481$ & $7.330161$ & $0.081771$ \\
\end{tabular*}
\end{ruledtabular}
\end{table*}

\begin{table*}[hbt!]
\caption{\label{tbl:FDFreqsB} Frequencies $f$ (kHz) and damping times $\tau$ (ms) of the first few modes of Eckart stars with EoS B \eqref{eq:EoS}, viscosities $\zeta_{c} = 2.7\,\hat{\zeta}\times10^{31}\viscgs$ and central density $\epsilon_{c}=4.5\times10^{15}\epscgs$ computed in the frequency domain. The perfect fluid frequencies agree to within $\lesssim 1\%$ with Table~A.$19$ in Ref.~\cite{Kokkotas:2000up}.}
\begin{ruledtabular}
\begin{tabular*}{\textwidth}{@{\extracolsep{\fill}}lcccccc}
& \multicolumn{2}{c}{$n=0$} & \multicolumn{2}{c}{$n=1$} & \multicolumn{2}{c}{$n=2$} \\
\cline{2-3} \cline{4-5} \cline{6-7} 
\multicolumn{1}{c}{$\hat{\zeta}$} & \multicolumn{1}{c}{$f$} & \multicolumn{1}{c}{$\tau$} & \multicolumn{1}{c}{$f$} & \multicolumn{1}{c}{$\tau$} & \multicolumn{1}{c}{$f$} & \multicolumn{1}{c}{$\tau$} \\
\hline
$0$ & $0.845772$ & $\infty$ & $7.593629$ & $\infty$ &$11.587969$ & $\infty$ \\
$0.01$ & $0.845745$ & $23.228672$ & $7.593605$ & $7.490563$ & $11.587910$ & $3.795469$ \\
$0.1$ & $0.843009$ & $2.322780$ & $7.591216$ & $0.749039$ & $11.582061$ & $0.379546$ \\
$0.5$ & $0.773598$ & $0.464124$ & $7.533134$ & $0.149726$ & $11.439473$ & $0.075906$ \\
$1.0$ & $0.495014$ & $0.231332$ & $7.349717$ & $0.074736$ & $10.983109$ & $0.037957$ \\
$2.0$ & $0.0$ & $0.545966$ & $0.0$ & $0.063377$ & $6.583406$ & $0.037134$ \\
\end{tabular*}
\end{ruledtabular}
\end{table*}

\section{Constrained BDNK Equations of Motion}\label{app:BDNK}
In this appendix, we describe the constrained system of equations governing the evolution of BDNK stars that we solve numerically in the time domain. This system is obtained from the $tt$, $tr$, $rr$ and $\theta\theta$ Einstein equations in addition to the $t$ and $r$ energy-momentum conservation equations. 

First, we solve the $rr$ Einstein equation for $\partial_{r}\delta\nu$ which eliminates $\delta\nu$ from the system. All time derivatives of $\delta\lambda$ can be eliminated by solving the $tt$ Einstein equation for $\partial_{t}\delta\lambda$, leaving three second-order equations and one first-order equation. To obtain wave equations for $\delta{u}$ and $\delta\epsilon$ with $\partial_{r}^{2}\delta\lambda$ eliminated, we take the following combinations
\begin{subequations}
    \begin{align}
        \textsf{B}_{1}(\mathrm{Eq.}\,2)+\textsf{B}_{2}(\mathrm{Eq.}\,3)+\textsf{B}_{3}(\mathrm{Eq.}\,4),\\
        \textsf{B}_{4}\left[\partial_{r}(\mathrm{Eq.}\,1)-\mathrm{Eq.}\,2\right]+\textsf{B}_{5}(\mathrm{Eq.}\,4),
    \end{align}\label{eq:app:BDNKTransformation}
\end{subequations}
where Eqs.~$1$-$4$ correspond, respectively, to the $tr$ and $\theta\theta$ Einstein and $t$ and $r$ energy-momentum conservation equations. The coefficients take the form
\begin{subequations}
    \begin{align}
        \textsf{B}_{1}&=\tau_{e} (p+\epsilon),\\
        \textsf{B}_{2}&=-e^{-\nu/2},\\
        \textsf{B}_{3}&=-4 \pi  r \tau_{e} (p+\epsilon),\\
        \textsf{B}_{4}&=4 \pi  r (p+\epsilon) (\tau_{e}+\tau_{q}) \left[3 \zeta+4 \eta-3 \tau_{p} (p+\epsilon)\right],\\
        \textsf{B}_{5}&=e^{-\lambda} \left[8 \pi  \tau_{e} (p+\epsilon) \left(3   e^{\nu/2}-2 \pi  r^2 e^{\lambda} (-3 \tau_{p} (p+\epsilon)+3 \zeta+4 \eta) \left(8 \pi    e^{\nu/2} \tau_{q} (p+\epsilon)-1\right)\right)+3\right].
    \end{align}
\end{subequations}
The constrained system we solve consists of the first-order $tr$ Einstein equation and the second-order equations resulting from \eqref{eq:app:BDNKTransformation}. It takes the form
\begin{align}
    \begin{pmatrix} 
     \bullet & \bullet & 0 \\ 
     \bullet & \bullet & 0 \\ 
     0 & 0 & 0 
    \end{pmatrix} \partial_{t}^{2}\mathbf{U} 
    &+ 
    \begin{pmatrix} 
     \bullet & \bullet & 0 \\ 
    \bullet & \bullet & 0 \\ 
    0 & 0 & 0 
    \end{pmatrix} \partial_{t}\partial_{r}\mathbf{U} 
    + 
    \begin{pmatrix} 
    \bullet & \bullet & 0 \\ 
    \bullet & \bullet & 0\\ 
    0 & 0 & 0
    \end{pmatrix} \partial_{r}^{2}\mathbf{U}
    + 
    \begin{pmatrix} 
    \bullet & \bullet & 0 \\ 
    \bullet & \bullet & 0 \\ 
    \bullet & \bullet & 0 
    \end{pmatrix} \partial_{t}\mathbf{U}
    + 
    \begin{pmatrix} 
    \bullet & \bullet & \bullet \\ 
    \bullet & \bullet & \bullet \\ 
    \bullet & \bullet & \bullet 
    \end{pmatrix} \partial_{r}\mathbf{U}
    + 
    \begin{pmatrix} 
    \bullet & \bullet & \bullet \\ 
    \bullet & \bullet & \bullet \\ 
    \bullet & \bullet & \bullet 
    \end{pmatrix} \mathbf{U} = 0, \label{eq:FullWaveSystem_Bullets}
\end{align}
where $\mathbf{U}(t,r)\equiv[\delta{u},\delta\epsilon,\delta\lambda]^{\rm{T}}$ and a $\bullet$ denotes a function of the background TOV solution. The coefficients in these equations are quite long, with a total leaf count of $\sim 10^{4}$ in \texttt{Mathematica}. We have consequently not written them here explicitly and instead provide them in accompanying \texttt{Mathematica} files.

\end{document}